\documentclass[a4paper,fleqn]{cas-dc}
\usepackage[authoryear]{natbib}

\usepackage{aas}

\newcommand{\ks}{km~s$^{-1}$}
\newcommand{\ms}{M$_{\odot}$}
\newcommand{\rs}{R$_{\odot}$}
\newcommand{\ls}{L$_{\odot}$}
\newcommand{\oc}{$O\!-\!C$}

\newcommand{\ond}{Ond\v{r}ejov}

\newcommand{\valmez}{Vala\v{s}sk\'e Mezi\-\v{r}\'{\i}\v{c}\'{\i}}
\newcommand{\hol}{Holovousy}
\newcommand{\n}{N103}
\newcommand{\T}{{\sc Tess}}

\newcommand{\mj}{M$_{\rm Jup}$}
\newcommand{\obj}{2M0410} 

\begin{document}
\let\WriteBookmarks\relax
\def\floatpagepagefraction{1}
\def\textpagefraction{.001}

\shorttitle{A photometric study of NSVS~01031772 and 2MASS J04100497+2931023}

\shortauthors{M. Wolf et al.}

\title [mode = title]{TESS light curves and surface activity in two low-mass eclipsing binaries: NSVS~01031772 and 2MASS~J04100497+2931023}                  

\author[1]{Marek Wolf}[orcid=0000-0002-4387-6358]
\cormark[1]
\ead{marek.wolf@mff.cuni.cz}
\credit{Methodology, Analysis of the data, Writing, review \& editing}

\affiliation[1]{organization={Astronomical Institute, Faculty of Mathematics and Physics, Charles University},
    addressline={V Hole\v{s}ovi\v{c}k{\'a}ch 2}, 
    city={Praha 8},
    postcode={CZ-180 00}, 
    country={Czech Republic} }   
\cortext[cor1]{Corresponding author}

\affiliation[2]{organization={ \valmez\ Observatory},
    addressline={ Vset\'{\i}nsk\'a 78}, 
    postcode={CZ-757~01},
    city={\valmez}, 
    country={Czech Republic} }      

\affiliation[3]{organization={Czech Astronomical Society, Variable Star and    Exoplanet  Section},
    addressline={Fričova 298}, city={\ond}, 
    postcode={CZ-251 65},
    country={Czech Republic} }

\affiliation[4]{organization={Research Centre for Theoretical Physics and Astrophysics, Institute of Physics, Silesian University},
    addressline={Bezru\v{c}ovo n\'am. 13}, 
    city={Opava}, 
     postcode={746 01}, 
    country={Czech Republic}  }    

\affiliation[5]{organization={Astronomical Institute, Czech Academy of Sciences},
    addressline={Fričova 298}, city={\ond}, 
     postcode={CZ-251 65}, 
    country={Czech Republic}   }       


\author[2,3]{Ladislav \v{S}melcer}[]
\credit{Photometric observations, Analysis of the data}

\author[1,3,4,5]{Hana Kučáková}[orcid=0000-0002-1330-1318]
\credit{Photometric observations}

\author[1]{Al\v zb\v eta Opli\v{s}tilov\'a}[orcid=0000-0002-3985-4463]
\credit{Numerical simulations}

\author[1]{Petr Zasche}[orcid=0000-0001-9383-7704]
\credit{Photomeric observations, Analysis of the data}

\author[6]{František B\'ilek} []
\credit{Photometric observations}


\author[7]{Andrej Mudray}[]
\credit{Photometric observations}

\author[8]{Tom\'a\v{s} Hynek}[]
\credit{Photometric observations}

\author[9]{Miloslav Zejda}[orcid=0000-0001-6231-3350]
\credit{Photometric observation, Analysis of the data}

\author[5]{Kamil Hornoch} [orcid=0000-0002-0835-225X]
\credit{Photometric observations}

\affiliation[6]{organization={Private Observatory},
    addressline={Trocnovská 1188}, 
    city={Trhové Sviny}, 
    postcode={CZ-374 01}, 
    country={Czech Republic} }

\affiliation[7]{organization={Private Observatory},
    addressline={Chloumky 31}, 
    city={Holovousy}, 
    postcode={CZ-508 01}, 
    country={Czech Republic}
    }       

\affiliation[8]{organization={DARK SKY Beskydy},
    addressline={Park tmav\'e oblohy}, 
    country={Czech Republic}
    }       

\affiliation[9]{organization={Institute of Theoretical Physics and Astrophysics,  Masaryk    University},
     addressline={Kotl\'a\v{r}sk\'a 2}, city={Brno}, 
      postcode={CZ-611 37}, 
     country={Czech Republic}    }

\begin{abstract}
The \textit{VRIC} light curves were regularly measured for two eclipsing binaries, NSVS~01031772 and 2MASS J04100497+2931023 as part of our long-term observational project to study low-mass eclipsing binaries with a short orbital period and surface activity.
The solution of the \T\ light curves in {\sc Phoebe} results in a detached configuration. 
Absolute parameters of all components were improved:
for \n:  $M_1 = 0.5475 \pm 0.0035$~\ms, $R_1 = 0.5297 \pm 0.0035$~\rs, 
          $M_2 = 0.5038 \pm 0.0040$~\ms, $R_2 = 0.5217 \pm 0.0035$~\rs, 
for \obj: $M_1 = 0.639 \pm 0.045$~\ms, $R_1 = 0.655 \pm 0.035$~\rs, 
          $M_2 = 0.609 \pm 0.045$~\ms, $R_2 = 0.631 \pm 0.035$~\rs, 
where the temperatures of the primary components were adopted according to previous studies. The spectral type of the primary components was confirmed to be M4 and K5, 
and the mass ratio was derived as $q = 0.920 \pm 0.003$ or $0.952 \pm 0.010$, respectively. 
We propose the presence of a third body in these systems: in the case of \n, a companion with a minimal mass of 50~M$_{\rm Jup}$, orbiting the eclipsing pair with a period of about 19~years, and in \obj\ a third body with a minimal mass of about 0.1 \ms\ and a short orbital period of about 2.1 years.
For \n, the hierarchical structure (2+1)+1 of a possible quadruple system was
tested, but its stability was not proven. 
The characteristics and statistics of the flare events and dark regions on the surface of the components were estimated on the basis of the \T\ and our own data. 
For \n, a mean frequency of flares of one per 40~hours was determined. In the case of \obj, practically no flares were detected.
\end{abstract}


\begin{keywords}
binaries: close \sep 
binaries: eclipsing \sep
binaries: low-mass \sep 
stars: activity \sep
stars: fundamental parameters \sep 
stars: individual: NSVS~1031772 \sep  
stars: individual: 2MASS~J04100497+2931023
\end{keywords}

\maketitle

\section{Introduction}
\label{sec:intro}

The most frequent stars in our Galaxy are late-type and low-mass stars, red dwarfs with masses below 1.0 \ms. Current observations of low-mass stars show a discrepancy between the estimated and modelled parameters, where the models give 10--15\% lower radii and higher temperatures than the observations \citep{2000ARA&A..38..337C, 2010ASPC..435..141M, 2015ApJ...804...64M, 2018MNRAS.476.5253C, 2022MNRAS.515.1416C}.
Many eclipsing binaries (EB) display a periodic variation of their mid-eclipse times
caused by various phenomena. One cause is the light-time effect (LITE) associated with the presence of a nearby third body orbiting the eclipsing pair \citep{1952ApJ...116..211I, 1959AJ.....64..149I}. The second alternative is magnetically induced gravitational modulation caused by an active star in the system \citep{1992ApJ...385..621A, 2020MNRAS.491.1820L}. The fundamental properties of late-type stars in EB were summarized by \cite{2022Galax..10...98M}, who confirmed the above-mentioned discrepancy and emphasized the importance of low-mass stars to the precision of our stellar evolution models. Thus, the sample of well-studied low-mass binaries needs to be increased. 

Similarly to our Sun, low-mass stars are also affected by chromospheric activity caused by a strong magnetic field. This variable activity has been frequently observed as surface spots or flares, which are sudden brightening events driven by the magnetic field. 
Spots on the surface play an important role in determining precise physical parameters, especially their radii and temperatures.
For the first time, spot activity in stars was announced by \cite{1950AJ.....55...69K}, who detected sinusoidal-like variations in the light curve of the eclipsing binary YY~Gem. 
It is therefore of great importance to monitor various manifestations of stellar activity in individual systems over the long term. 

 Moreover, the possible multiplicity of low-mass systems is relatively rare \citep{2013ARA&A..51..269D}, and therefore, it is very important to increase the number 
of possible candidates for better statistics. 
In this paper, we report long-term photometric and mid-eclipse time monitoring of NSVS~01031772 and 2M0410, two relatively well-known low-mass eclipsing binaries that consist of similar M4 and K5 dwarf components. 

 Our previous studies of active low-mass eclipsing binaries were presented in \cite{2017MNRAS.466.2542S, 2023MNRAS.520..353S}.
This paper is organized as follows. 
The next Section~\ref{objects} contains all known information about these objects.
In Section~\ref{sec:obs}, we present the observational data used in the analysis. 
The orbital period study and the light-time effect are described in Section~\ref{sec:period}, where the stability of the possibly quadruple system \n\
is also discussed. 
The photometric solution and absolute dimensions are given in Section~\ref{sec:curve}.
The statistics of the flare activity of \n\ and surface spots on \obj\ are given in Section~\ref{sec:flare}.  
Our conclusions are briefly summarized in Section~\ref{sec:concl}.

\section{Objects}
\label{objects}
Two similar northern low-mass eclipsing binaries of late spectral types were included in this study. They have been intensively observed in the past, and it is possible to derive new or improved parameters for their components. 

\subsection{NSVS~01031772}
\label{n103info}

The eclipsing binary NSVS~01031772 (also  2MASS J13453489+7923482, TIC~155657581,
hereafter \n, $V = 13.4$ mag)
is a nearby detached low-mass binary with a short orbital period of about 9~hours. 
It was first announced as FBS~1345+796 in the First Byurakan Spectral Sky Survey \citep{1995Ap.....38..195A}.
This relatively well-studied binary was also included in the NSVS survey \citep{2004AJ....127.2436W}. The first photometric study was presented by \cite{2005IAPPP.101...38M}, where the orbital period of 0.368154(4) day 
was given. 
The only comprehensive photometric and spectroscopic study by \cite{2006astro.ph.10225L} 
derived very precise absolute parameters:
$M_1 = 0.5428 \pm 0.0027$~\ms, $R_1 = 0.5260 \pm 0.0028$~\rs, 
$M_2 = 0.4982 \pm 0.0025$~\ms, $R_2 = 0.5088 \pm 0.0030$~\rs, 
and the distance $d = 60.6\pm 0.7$~pc, which is in good agreement with the current 
{\sc Gaia} estimation (60.7~pc).
Similarly to other low-mass binaries studied, they conclude that the radius of each star exceeds the best evolutionary model predictions by about 8.5~\% on average.
They derived a more precise light ephemeris:

\begin{center}
   Pri.Min. = HJD~24~53456.6796(2) + $0.3681414(3) \cdot E$.
\end{center}

\noindent
No flares were recorded in their photometric data sets.
The space velocities of the system suggest that it is a main sequence binary with a metallicity that is approximately solar. The two components in this binary are located in a region of the mass-radius diagram where no accurate observational data were previously available. 

Later, a possible light-time effect was proposed, and the next analysis of the light curve was given in \cite{2012IAUS..282..490W}.
They predicted the presence of a third body in this system orbiting with a period of at least 11~yr. The {\it V,R} light curves obtained at the \ond\ observatory were solved independently in {\sc Phoebe}, and results similar to \cite{2006astro.ph.10225L} were found. 
No eruption activity was recorded, but a dark region on the primary component 
(temperature factor $T_f$ = 0.95, latitude 90~deg, longitude 180~deg and radius 25~deg) was used in the solution of the light curve.

Since 2013, \n\ has been systematically measured photometrically by a group of amateur observers from the Czech Astronomical Society with the aim of monitoring the eruptive activity of its components. The first results were published in \cite{2016OEJV..175....1S}.
They detected five eruptions between 2013 and 2015 and concluded that the rate of eruptions appears to be randomly distributed over time, and no periodicity was found. The first estimates of the energy released were made $E \simeq 10^{27} - 10^{28}$ J.

\n\ is included in many other catalogs: 
{\it Nearby and high proper motion star} of \cite{2011AJ....142..138L},
{\it HST/FGS Trigonometric Parallaxes of M-dwarf Eclipsing Binaries} by \cite{2020PASP..132e4201V}, and 
as a high proper motion star PM~J13455+7923 in \cite{2020yCat.1350....0G} catalog.
\n\ is also listed in the ROSAT All-Sky Bright Source Catalog \citep{1999A&A...349..389V}
as X-ray source 1RXS J134540.6+792332. 
\n\ is included in the ASAS-SN database \citep{2014ApJ...788...48S} as an object 
AP~9212951. Unfortunately, the large scatter of these data does not allow us to determine a reliable solution for the light curve or the precise mid-eclipse times.

\begin{table}
\begin{center}
\caption{ {\sc Gaia} DR3 astrometric and photometric data on \n\ and \obj.}
\label{tg}
\begin{tabular}{ccc}
\hline\hline\noalign{\smallskip}
 Parameter              &  \n                &   \obj     \\
\hline\noalign{\smallskip}
$\alpha_{2000}$ [h m s] &  13 45 34.871      &  04 10 04.970   \\
$\delta_{2000}$ [d m s] &  +79 23 48.29      & +29 31 02.258   \\
pm $\alpha$ [mas/yr]    & --106.85 $\pm$ 0.07  & -29.663  $\pm$ 0.061 \\
pm $\delta$ [mas/yr]    &   59.023 $\pm$ 0.06  & -20.819  $\pm$ 0.038 \\ 
parallax $\pi$ [mas]    &   16.572 $\pm$ 0.074 &   6.485  $\pm$ 0.050 \\ 	
{\sc Gaia} RUWE         &   1.506              &    1.498  \\
\hline\noalign{\smallskip}
$B$ [mag]               &   --                &  16.635    \\
$V$ [mag]               & 13.363  $\pm$ 0.038 &  --  \\
$R$ [mag]               & 13.074  $\pm$ 0.038 &  13.750    \\
$G$ [mag]               & 12.363  $\pm$ 0.003 &  13.8393 $\pm$ 0.0021 \\
$BP$ [mag]              & 13.674  $\pm$ 0.009 &  15.0633 $\pm$ 0.0141 \\
$RP$ [mag]              & 11.223  $\pm$ 0.006 &  12.7594 $\pm$ 0.0059 \\
$J$ [mag]               & 9.692   $\pm$ 0.022 &  11.131  $\pm$ 0.021  \\
$H$ [mag]               & 9.021   $\pm$ 0.016 &  10.375  $\pm$ 0.020  \\
$K$ [mag]               & 8.778   $\pm$ 0.016 &  10.133  $\pm$ 0.017  \\
\hline
\end{tabular}
\end{center}
\end{table}

\subsection{2MASS J04100497+2931023}
\label{2minfo}

 The eclipsing binary 2MASS~J04100497+2931023
(also T-Tau0-07388, TIC~56395514, 
hereafter \obj, $G_{\rm max} = 13.84 $ mag) 
is a lesser-known northern object with a short orbital period of about 14~h.  
Its variability was discovered by \cite{2008AJ....135..850D} with a relatively precise
orbital period of $P = 0.6078457$ days.
The only photometric and spectroscopic study of \obj\ was presented by \cite{2021RAA....21..115M}, who derived relatively precise masses of the components with the more massive secondary: 
$M_1 = 0.587 \pm0.006$ \ms, 
$M_2 = 1.049 \pm 0.0113$ \ms. 
They also obtained the orbital period of 0.60783(5) day.
\obj\ is included in the catalog of eclipsing binaries observed by LAMOST
\citep{2018ApJS..235....5Q}, where the mean effective temperature of 5365~K 
is given
and two values of radial velocities are listed: --17.0 and +10.4~km/s. 
The following linear ephemeris was proposed in the VSX index~\footnote{\url{https://www.aavso.org/vsx/}} for current use:
$$ {\rm Pri.Min. = HJD\;24\;53736.925 + 0.6078457} \cdot E. $$  
 
\noindent
The {\sc Gaia} DR3 astrometric and photometric data on \n\ and \obj\ are summarized in Table~\ref{tg} \citep{2022yCat.1355....0G}. 
The RUWE parameter is the renormalized unit weight error and serves as an indicator of the binarity of the object \citep{2021A&A...649A...2L}. RUWE > 1.4 is actually a reliable indicator of the presence of a close companion \citep{2021ApJ...907L..33S}.
To our knowledge, no modern photometric or period analyzes of these short-period and low-mass eclipsing binaries exist so far.

\section{Observations}
\label{sec:obs}

In this section, we describe the properties and reduction process
of our photometric observations of \n\ and \obj. We present our long-term collection of ground-based photometry (\ref{sec:ground}), extensive {\sc Tess} photometry (\ref{sec:tess}), and older optical spectra from the KPNO and Keck Observatories (\ref{sec:Spec}). 


\subsection{Ground-based photometry}
\label{sec:ground}

Since 2007, the time-resolved CCD photometry of \n\ and \obj\ mostly during eclipses, has been regularly obtained at several observatories in the Czech Republic. 
The focused photometric campaign on \n\ was initiated in July 2013 to determine the eruption activity and flare frequency to derive their characterization, and a relatively large amount of photometric data was obtained. In chronological order:

\begin{itemize}
\item Since 2007, semi-regular observations of both objects have been made at \ond\ observatory, Czech Republic, where the Mayer 0.65-m ($f/3.6$) reflecting telescope with the CCD camera MII G2-3200 and $VR$ photometric filters were used. Typical exposure times of 30 to 60 seconds and binning $2 \times 2$ were applied during remote access.

\item Most of the observations in our campaign were systematically obtained at the Observatory \valmez\ in the Czech Republic~\footnote{\url{https://www.astrovm.cz}}, from July 2013 to {May 2026}. Four different telescopes, CCD cameras, and various filters were used each clear night:
Celestron SCT 280/1765, MII G2-4000 camera with filter $R$, 
Sky-Watcher NWT 254/1200, QHY~174 CCD camera without a filter, 
Celestron SCT 355/2460, MII G2-1600 camera with filters $VI$, and
Newton 500/1900, MII C4-16000 camera with Sloan filters g´r´.
 
\item Long-term photometric monitoring and a substantial part of our data set for \n\ were obtained at the private observatory of F.B. at Trhové Sviny observatory, South Bohemia, Czech Republic, where a Newtonian 200/860 telescope and CCD camera Atik~314L+ were used. Exposures were taken without the filter, with a typical exposure time of 90~s. Since 2015, almost 20~000 exposures have been obtained over 70~nights, corresponding to 430~hours of systematic monitoring. 

\item Additional photometry of \n\ was obtained at the former Masaryk University Observatory in Brno (MUO), Czech Republic, during several separate epochs: in Oct/Dec~2009, Feb~2017 -- Apr~2019, and Feb/Jul~2023. 
The 0.60-m Newtonian reflecting telescope with the CCD camera SBIG~ST8, G2-4000 or G4-16000, and {\it VR} photometric filters was included. 


\item   Important photometry of \obj\ was obtained at the private observatory of A.M. in \hol\ near Ho\v{r}ice, Czech Republic: the SCT 0.406-m ($f/6.3$) telescope, reducer, and coma corrector Starizona 0.63x, CMOS camera ZWO ASI071. The clear filter and a 60-second exposure time were used during the 2024-2025 season. 

\item  Finally, CCD photometry of \obj\ was secured at Beskydy Dark Sky Observatory, Czech Republic~\footnote{\url{https://www.darkskybeskydy.cz}}: the 40-cm Ritchey-Chrétien telescope, CCD camera MII~C3-Pro mono with Baader LRGB Halfa filters in remote access.  \\

\end{itemize}

\noindent
The CCD observations were reduced in a standard way. 
The {\sc C-Munipack}\footnote{Package of software utilities 
for reducing astronomy CCD images, current version 2.1.31, 
available at \url{http://c-munipack.sourceforge.net/}},
an aperture photometry software, was routinely used for our time series photometry.
In \ond, the {\sc Aphot}, synthetic aperture photometry and astrometry software, was used for time series.
Time series were constructed by computing the magnitude difference between the variable and a nearby comparison and check star; the heliocentric correction was applied. 
The uncertainties of the photometric measurements at smaller telescopes were always about 0.01 -- 0.02~mag.
The computers at our telescopes are synchronized using different time servers provided  for example, by {\sc Cesnet} Time Services~\footnote{CESNET: \url{https://www.cesnet.cz/en}} every two minutes. 
These corrections are usually on the order of $10^{-3}$ seconds.
Together, it was obtained at all observatories over $2 \times 10^5$ frames 
in the $V, R, I$ filters or without the filter mentioned in the text as $C$ - clear.

\subsection{TESS photometry}
\label{sec:tess}

\begin{table}
\centering
\caption { The \T\ visibility of \n\  and \obj\ 
and sectors used for light curve analyses (LC) and for mid-eclipse 
time determination (Min).}
\label{tess_sec}
\begin{tabular}{ccccc}
\hline\hline\noalign{\smallskip}
 Sector  &  Start date  & Exposure   &  Used for  \\
  No.     & YYYY-MM-DD  & time [sec] &  solution \\
\hline\noalign{\smallskip}
\multicolumn{4}{c}{ \n}  \\
    14 & 2019-07-18  &  120 &  Min \\
    20 & 2019-12-25  &  120 &  Min \\
    21 & 2020-01-21  &  120 &  Min \\
    26 & 2020-06-09  &  120 &  Min \\
    40 & 2021-06-25  &  120 &  Min \\   
    41 & 2021-07-24  &  120 &  Min \\
    47 & 2021-12-31  &  120 &  Min \\
    48 & 2022-01-28  &  120 &  LC, Min \\
    53 & 2022-06-13  &  120 &  LC, Min \\
    60 & 2022-12-23  &  120 &  LC, Min \\
    73 & 2023-12-07  &  120 &  Min \\
    74 & 2024-01-03  &  120 &  LC, Min \\
    75 & 2024-01-30  &  120 &  Min \\  
\hline\noalign{\smallskip}
\multicolumn{4}{c}{ \obj}   \\
    43   & 2021-09-16  &  600  &  Min \\
    44   & 2021-10-12  &  600  &  LC, Min \\
    59   & 2022-11-26  &  200  &  LC, Min \\
    70   & 2023-09-20  &  200  &  LC, Min \\ 
    86   & 2024-11-21  &  200  &  LC, Min \\
\hline
\end{tabular}
\end{table}

As an object with high northern declination (Dec. $\simeq$ +80 deg), \n\ was measured many times by the {\it Transiting Exoplanet Survey Satellite} ({\sc Tess}, 
\cite{2015JATIS...1a4003R}).  
We used the MAST~\footnote{MAST: Barbara A. Mikulski Archive for Space Telescopes, \\ \url{https://mast.stsci.edu/portal/Mashup/Clients/Mast/Portal.html}} database to download the photometric time-series. 
All available {\sc Tess} Sectors of \n\ with a short exposure time of 120~s are listed in  Table~\ref{tess_sec}.
 A total of five {\sc TESS} sectors with exposure times of 200 or 600 seconds were used for \obj.
These high-quality light curves allowed us to determine the precise mid-eclipse times as well as to model the system.

The new times of primary and secondary minima and their uncertainties were generally determined by fitting the light curve with Gaussians or polynomials of the third or fourth order; we used the least-squares method. 
 From the numerous {\sc Tess} sectors, the new mid-eclipse times were derived only from the beginning, middle, and end of each sector. A higher number of minima is already redundant and will not substantially increase the accuracy of our results.
They are listed in Tables~\ref{t1}, \ref{tess} and \ref{2m04mintess}, where the epochs were calculated according to the improved linear ephemeris given in Table~\ref{t2}.
Because {\sc Tess} data are provided in the Barycentric Julian Date Dynamical Time (BJD$_{\rm TDB}$), all our times in Table~\ref{t1} were first transformed into this time scale using the NextAstro Hub \footnote{\url{https://arbiter.nextastro.org/toolkit/bjd-converter}}.

\subsection{Spectroscopy}
\label{sec:Spec}

Fortunately, time-compact high-dispersion spectroscopy and radial velocities of \n\ were obtained by \cite{2006astro.ph.10225L}. 
They collected a total of 108 spectra during two nights in May 2005 with the echelle spectrograph at the 4-m~Mayall telescope at the Kitt Peak National Observatory (KPNO). The wavelength coverage of each spectrum is 5700 -- 8160~\AA, with a resolution power of 18~750.
We used this set of precise radial velocities in our subsequent {\sc Phoebe} solution. 

\begin{figure}
\begin{center}
\includegraphics[width=\linewidth]{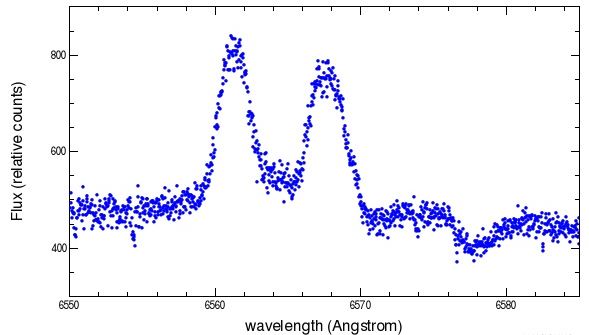}
\caption[ ]{The emission double-peaked hydrogen H$\alpha$ line profile 
of \n\ obtained at Keck Observatory in June 2011 as a strong indicator 
of chromospheric activity.} 
\label{keckspec}
\end{center}
\end{figure}

Two spectra of \n\ obtained on June 27--28, 2011, are available in the Keck Observatory Archive (KOA)\footnote{KOA Data Access Service - v18.5, \url{https://koa.ipac.caltech.edu/cgi-bin/KOA/nph-KOAlogin}}.
The HIRES spectrograph with exposure times of 180~s and 300~s was used. The whole calibrated spectrum covers the wavelengths 4450 -- 8910 \AA\ and clearly shows the double-peaked emission lines of the Balmer series of hydrogen (see Fig.~\ref{keckspec} for illustration). 
This emission is a clear manifestation of the chromospheric activity of both components.

 Concerning \obj, using {\sc Lamost} Medium Resolution Survey, a partially covered radial velocity curve was obtained by
\cite{2021RAA....21..115M} in January 2019. It was used in our subsequent solution in {\sc Phoebe}.

\section{Orbital period study}
\label{sec:period}

The period changes of \n\ were announced for the first time in  \cite{2012IAUS..282..490W}. 
To our knowledge, no period study has been presented for \obj\ so far. 
To confirm our previous finding, we continued to monitor eclipses to extend the \oc\ diagram and reveal the nature of long term period changes. Based on the current data set, extended by the next 10~years of continuous measurements, we propose in both cases triple systems, where the eclipsing pair orbits an additional body. 
Such a long-term project was possible only with the help of several advanced amateur observers using small telescopes and modern CCD techniques.

\subsection{Light-time effect}

\begin{figure}
\centering
\includegraphics[width=\columnwidth]{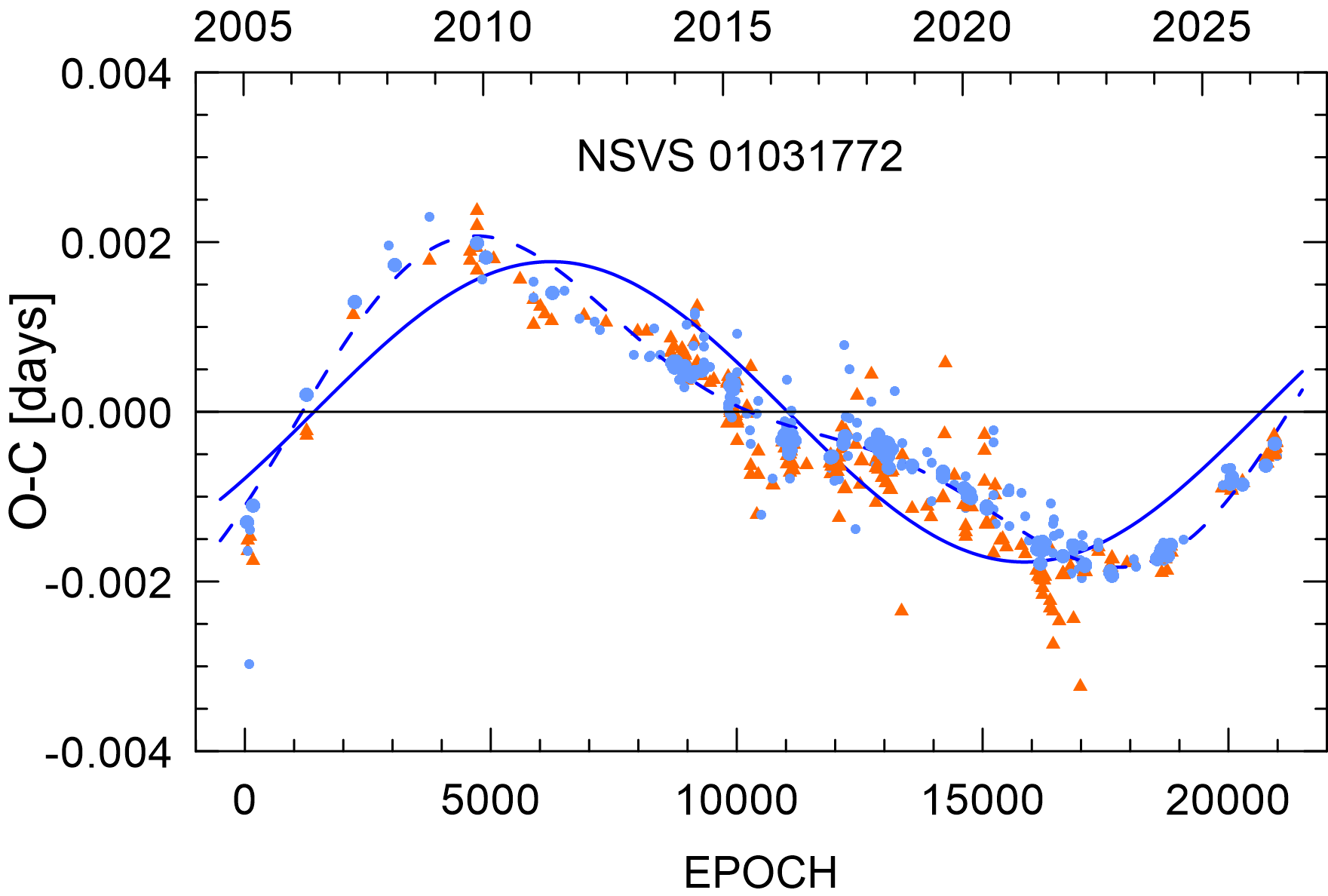}   
\includegraphics[width=\columnwidth]{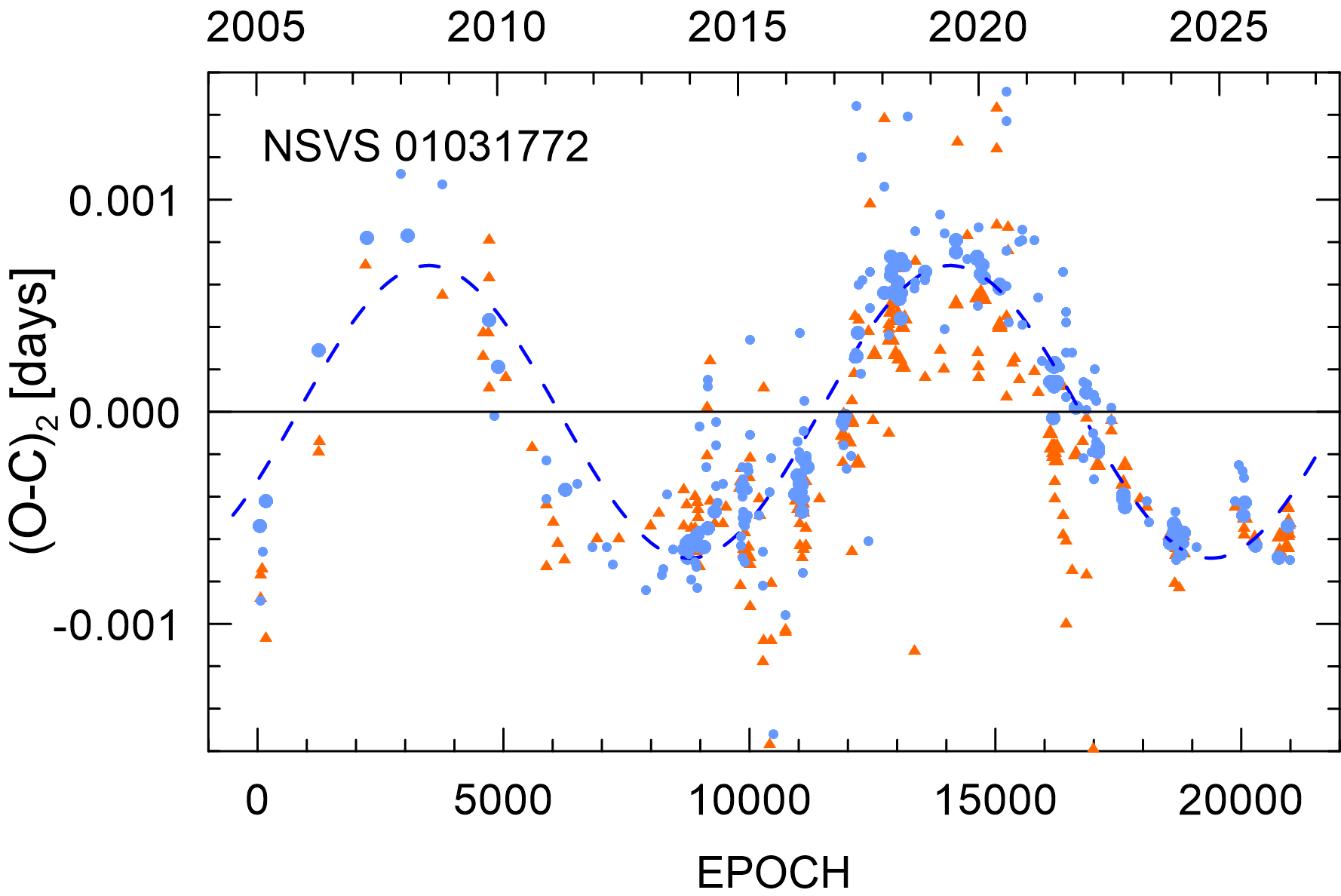}   
\caption[]{Top: The historical \oc\ diagram for \n\ since its discovery in 2005. The individual primary and secondary CCD minima are denoted by blue circles and red triangles, respectively. Larger symbols correspond to the precise CCD measurements, which were used in our calculation of the LITE.  The blue sinusoidal curve represents LITE with period of about 19~years and an semi-amplitude about 0.0018 days. The blue dotted curve is combination of LITE with a next variation which fit the minima better. 
Bottom: (\oc)$_2$ diagram after subtraction of LITE with the period of 19~years. 
The blue dotted curve denotes a next sinusoidal variation with a period of about 11~years and an amplitude of about 60 seconds caused probably by changes in magnetic field in components or an up-to now unknown phenomenon. The fourth body can be excluded, see Section~\ref{sec:stabil}.  }
\label{n103}
\end{figure}

\begin{figure}
\centering
\includegraphics[width=\columnwidth]{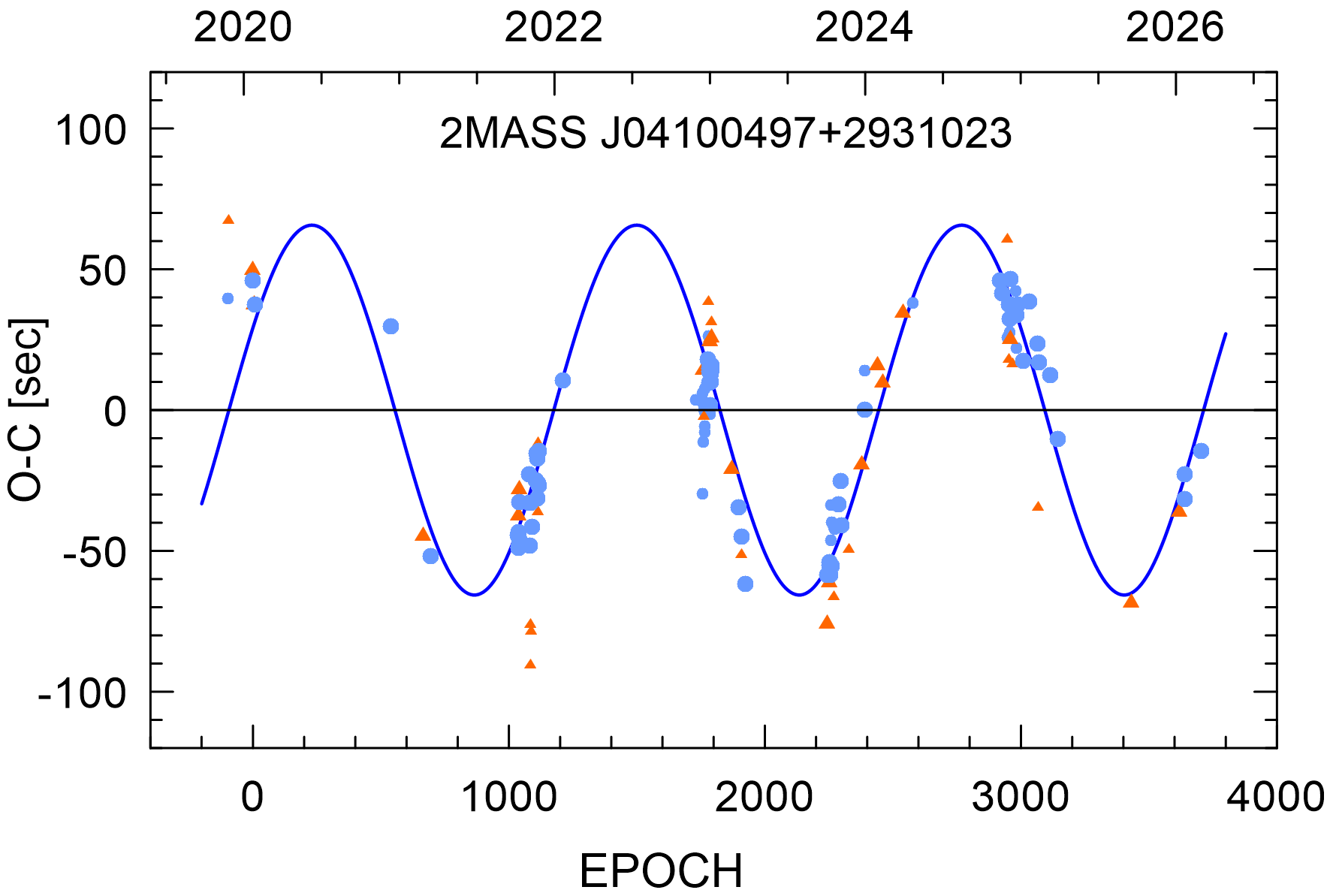}
\caption[ ]{The current \oc\ diagram for the times of minimum of 
\obj. The blue sinusoidal curve represents the LITE with a short period of about 2~years and a semi-amplitude of about 70~sec. The individual primary and secondary minima are denoted by blue circles and orange triangles, resp. 
\T\ minima are groups of points around epochs 1100, 1800, 2200 and 3000. 
Three orbits of a possible third body are covered. }
\label{2m04oc}
\end{figure}

The light-time effect (hereafter LITE) in both systems was traditionally analyzed using the \oc\ diagram (or ETV curve) with a sinusoidal shape.
As has been proven many times in the past, this relatively simple and effective technique is a powerful tool for identifying the multiplicity of stellar systems
 or even detecting brown dwarfs and extrasolar planets.
For a detailed description of the LITE analyzes, see some of the original papers of 
\cite{1952ApJ...116..211I, 1959AJ.....64..149I, 1973A&AS...12....1F, 1990BAICz..41..231M, 2005ASPC..335.....S, 2012AN....333..754P}. 
Concerning the light travel time caused by the third body, we refer the reader to our previous study of a similar low-mass system NSVS~7453183 \citep{2023MNRAS.520..353S}, where the relevant equations are provided. There are seven independent variables to be determined in this procedure:










\smallskip 
\noindent
$A_3$ -- the observed semi-amplitude of the light-time curves, \\
$P_3$ -- the orbital period of the third body, \\
$T_3$ -- the time of periastron, \\
$\omega_3$ -- the length of periastron, and \\
$e_3$ -- the eccentricity of the third-body orbit. \\
The linear ephemeris ($T_0, P_\mathrm{s}$) of the eclipsing pair is part of this set of  unknown parameters. 

\begin{table*}
\begin{center}
\caption{The LITE elements and the minimal masses of the possible third body in \n\ and \obj.}
\label{t2}
\begin{tabular}{ccccc}
\hline\hline
Parameter       & Unit  &    \n      &    \obj  \\
\hline\noalign{\smallskip}
$T_0$           & BJD   & 24~53456.68211 (3) & 24~58845.26951 (5)    \\
$P_s$           & day   & 0.368140108 (8)    & 0.60784724 (4)  \\
\hline\noalign{\smallskip}
$A_3$           & days  &     0.00177 (10) &  0.00076 (5)   \\
$A_3$           & sec   &     153 (9)      &  65.7 (4.3)       \\ 
$P_3$           & days  &     7090 (350)   &  771  (12)    \\
$P_3$           & years &     19.4 (1.0)   &  2.11 (0.03)    \\
$e_3$           &  --   &     0 (fixed)    &  0.035 (0.02)    \\
$\omega_3$      & deg   &     0            &  270.6 (2.5)   \\
$T_3$           & JD    &     24~53975 (150) & 24~58600 (10) \\
$f(M_3)$        & \ms   &     0.000 764    &  0.000 511     \\
$M_{\rm 3, min}$ & \ms  &     0.046        &  0.098    \\
$K_3$           & \ks   &     0.47         &  1.9      \\
$A_{\rm dyn}$   &  days &     0.000 03     &  0.000 016 \\   
\hline\noalign{\smallskip}
$\sum{w\ (O-C)^2}$ & day$^2$ &  1.63$\cdot10^{-5}$ & 6.8$\cdot10^{-6}$   \\ 
\hline
\end{tabular}
\end{center}
\end{table*}

The period analysis of \n\ and \obj\ was performed using all available mid-eclipse times found in the VarAstro database~\footnote {VarAstro, \url{http://var.astro.cz/en}}, primarily on our long-measured series of mid-eclipse times. In addition to those minima given in Tables~\ref{t1}, \ref{tess} and \ref{2m04mintess}, we used for \n\ the previous minimum times obtained by \cite{2006astro.ph.10225L}.
A total of 453~precise CCD times, including 210~secondary eclipses, were used for the analysis of \n. In the case of \obj, we collected 121~primary and secondary mid-eclipse times.  The least-squares method was applied, and the resulting LITE parameters and their internal fit errors are given in Table~\ref{t2} (in parentheses). 
The historical \oc\ diagram since the discovery of \n\ and \obj\ as variable stars is plotted in Fig.~\ref{n103} and Fig~\ref{2m04oc}. It is clearly visible that mid-eclipse times do not follow a simple linear ephemeris during the past years. 

The more complicated \oc\ diagram for \n\ led us first to a possible solution of two LITE caused by a third and fourth body orbiting in circular orbits. On the other hand, such a multiple system is unstable; see the following Section~\ref{sec:stabil}. We therefore consider the presence only of the third body. The (\oc)$_2$ diagram for \n\ after removing the 19-yr LITE period is shown in the bottom panel. It is clearly visible that the next sinusoidal shape of the \oc\ residuals has a period of about 11 years and an amplitude of 60 seconds. Due to the insufficiently covered LITE period of 19~years, we fixed $e_3=0$ in our final solution.
In the case of \obj, the third-body orbit is well-covered three times by our observations.


Assuming a coplanar orbit of the third body ($i_3 \simeq i$) and the total mass of the eclipsing pair $M_1 + M_2$ from our solution (see Tables~\ref{t3} and ~\ref{t4} below), we can estimate a lower limit for the mass of the third component, $M_{\rm 3, min}$ 
This value, as well as the mass functions $f(M_3)$, and
the amplitudes of the corresponding systemic radial velocities $K_3$ are also given in Table~\ref{t2}. 
 In the case of \n, a possible third component may be a brown dwarf with a minimal mass of 50 \mj. In \obj\ the companion could be a red dwarf M6-M7 with a minimal mass of about 0.1 \ms. 

We also tested the physical delay of the direct gravitational influence of the additional body on the motion of an eclipsing pair. The amplitude of the dynamical contribution of the third body, 
$A_{\rm dyn}$, is given by \cite{2016MNRAS.455.4136B}:

$$ A_{\rm dyn} = \frac{3}{8\pi} \frac{M_3}{M_1+M_2+M_3} \frac{P_s^2}{P_3} 
             \, \left(1-e^2_3\right)^{-3/2} $$

\smallskip
\noindent
and is also listed in Table~\ref{t2}. The value of $A_{\rm dyn}$ is on the order of seconds and is comparable  to the uncertainty in the estimation of the individual mid-eclipse time. Thus, physical delay cannot contribute significantly to the observed changes in \oc\ values.

\begin{figure}
\centering
\includegraphics[width=0.4\textwidth]{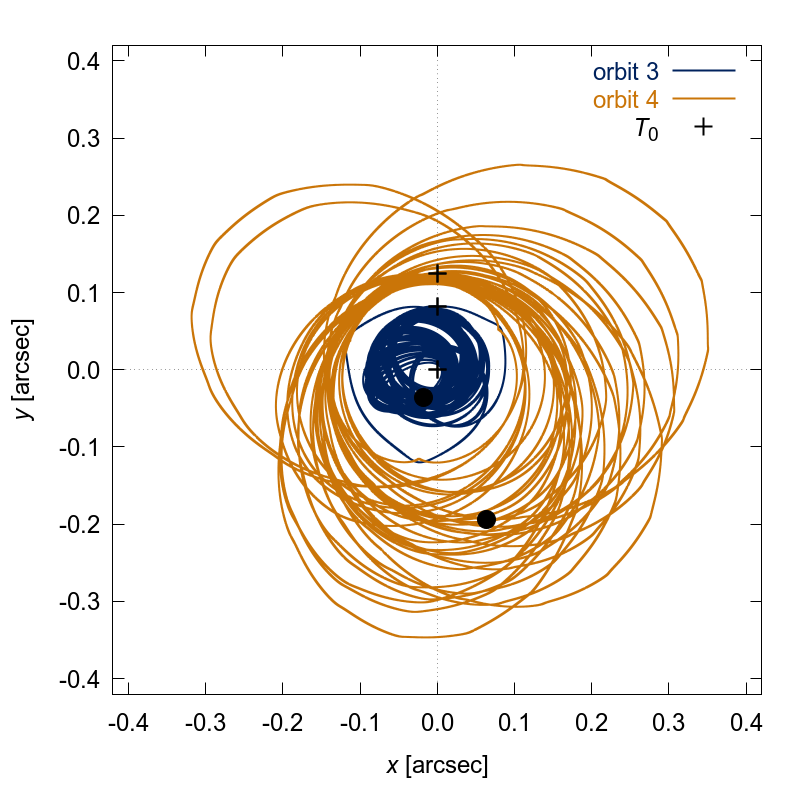}
\includegraphics[width=0.4\textwidth]{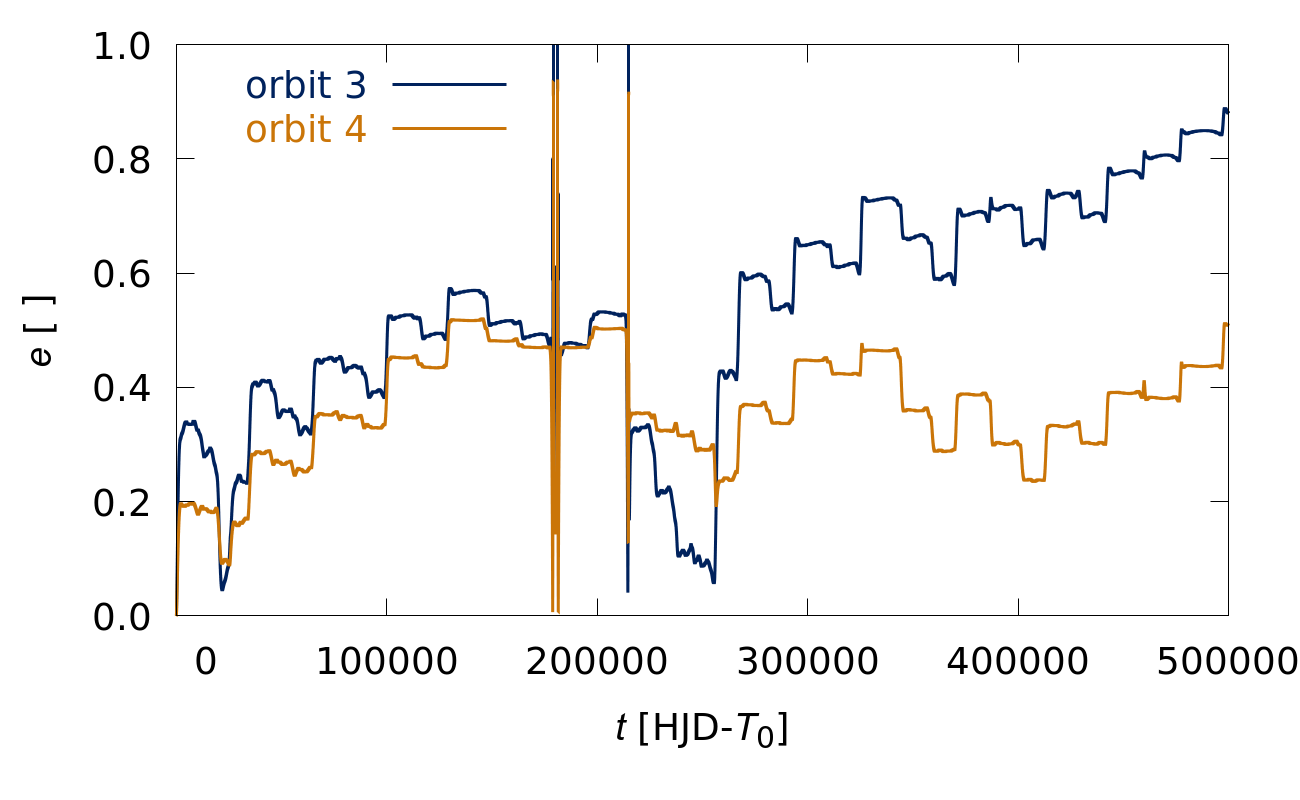}
\includegraphics[width=0.4\textwidth]{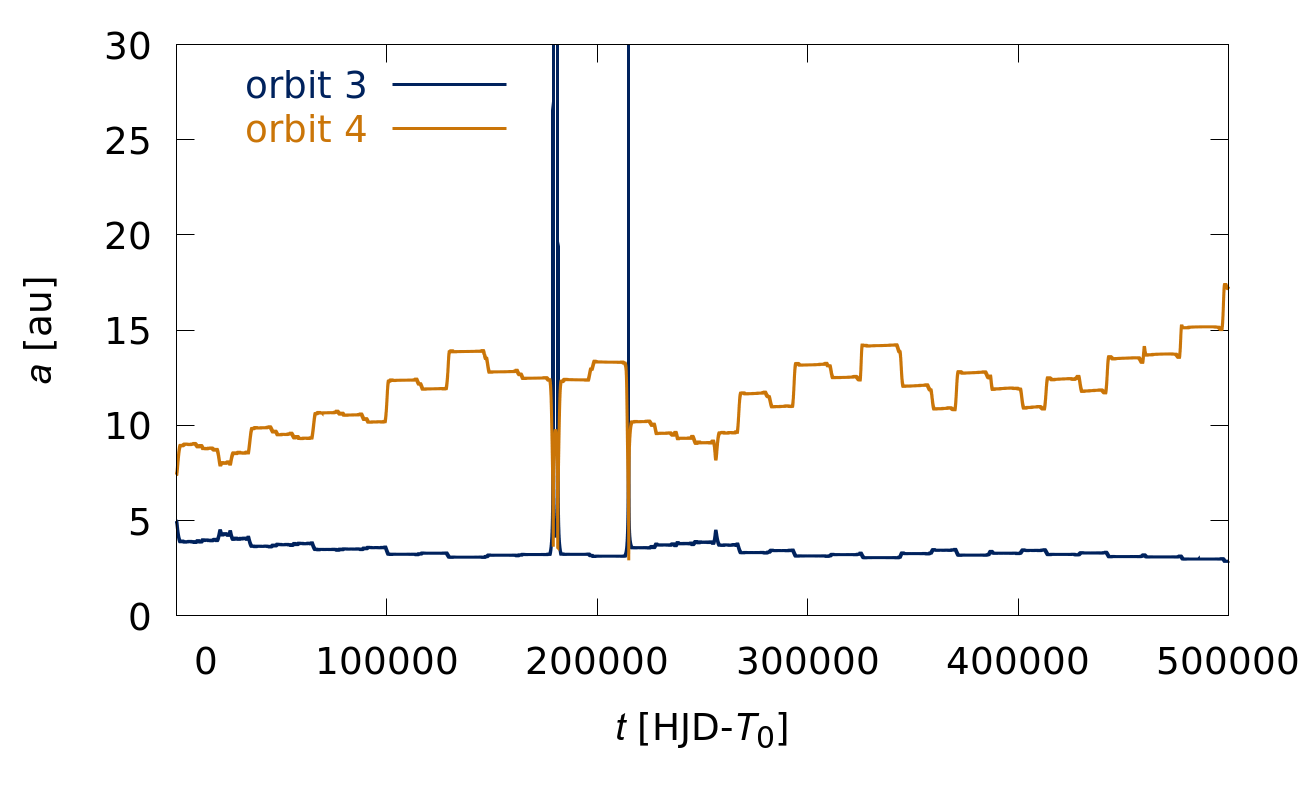}
\caption{
Top: Orbits of the possible third and fourth bodies in \n\ evolved over the subsequent 500\,000~days after $T_0$ denoted by crosses.
Middle: Evolution of eccentricity.
Bottom: Evolution of semi-major axis. 
The system exhibits large amplitude variations in orbital parameters.
At the beginning, we see angular momentum transfer from the inner orbit to the outer one.
The stepwise changes in $e$ and $a$ indicate energy exchange jumps and 
secular eccentricity pumping.
Sometimes, the eccentricities spike to near-unity, and the motion departs strongly from a Keplerian approximation.
The system undergoes secular eccentricity growth followed by a breakdown of the hierarchical configuration and chaotic scattering on very short timescales. 
}
\label{orbits}
\end{figure}

\subsection{Stability of N103 orbits}
\label{sec:stabil}

To investigate the dynamical stability of a possible quadruple system \n, we performed direct $N$‑body integrations of the system using the {\sc Xitau} code \citep{Broz_2017ApJS..230...19B,Broz_2021A&A...653A..56B,Broz_2022A&A...657A..76B,Broz_2022A&A...666A..24B}\footnote{\url{http://sirrah.troja.mff.cuni.cz/~mira/xitau/}}, which provides a high-precision Bulirsch--Stoer integration \citep{STOER1966} tailored to hierarchical multiple systems.
The time evolution of the orbital elements was analyzed using the {\tt xvpl2el} utility included in {\sc Xitau}. The utility converts Cartesian positions and velocities into osculating elements. This allowed us to monitor the temporal evolution of the orbital parameters of all components.
In the integrations, the close compact binary was treated as a single body. We evolved the orbits of the third and fourth components, assuming, as before for minimal mass estimate, coplanar configurations with aligned nodes (inclination $i = 0$, longitude of the ascending node $\Omega$ undefined, longitude of periastron $\varphi = 0$, true longitude $\lambda = 0$) and initially circular orbits ($e = 0$) with parameters as in Table~\ref{t2}. The system was integrated for 500\,000~d ($\approx 1370$~yr). 

We find the system to be dynamically unstable on short timescales. The first close encounter led to a rapid and significant change in the osculating orbital elements, with eccentricities increasing strongly (see Fig.~\ref{orbits} for the orbits and the eccentricity and semi-major axis evolutions). Within less than 2000~d, the eccentricity increased from 0 to $\approx 0.2$ for the fourth body and from 0 to $\approx 0.3$ for the third orbit. Correspondingly, the semi-major axis increased from 7.4~au to 8.9~au for the fourth orbit, while it decreased from 5.0~au to 3.9~au for the third orbit.
As the eccentricities grow, the radial ranges of the third and fourth orbits begin to overlap. This occurs when the periastron distance of the fourth orbit becomes comparable to or smaller than the apastron distance of the third orbit: 
$a_4 (1-e_4) \lesssim a_3(1+e_3)$.
 Once orbital overlap occurs, close encounters become possible and the system rapidly develops chaotic behaviour.
We verified this by integrating from nearly identical initial conditions, differing only by small perturbations in the initial eccentricity of $10^{-2}$, $10^{-3}$, $10^{-4}$, $10^{-5}$, $10^{-6}$ (a hallmark of chaotic dynamics). In addition to the coplanar configurations, we explored a range of initial inclinations; however, no long-term stable configurations were identified. The system appears to be too compact and not hierarchical enough.

\subsection{Magnetic activity}

A real alternative to the LITE and the third body solution in the case of cyclic ETV variations is the assumption of period modulation connected with the magnetic activity of stars. \cite{1992ApJ...385..621A} and \cite{2020MNRAS.491.1820L} proposed a model that explains the modulations of the orbital period as a consequence of changes in the magnetic activity of one of the components, or due to variations in the quadrupole moment $Q$. According to this model, the component that has a magnetic activity cycle can show a relative orbital period change of 

$$\frac {\Delta P}{P} = 2\pi \frac{O-C}{P_{\rm mod}} = -9\, \frac{\Delta Q} {M a^2}, $$

\noindent where $M$ is the mass of the active star and $a$ is the semi-major axis of the orbit. Usually, this mechanism cannot contribute significantly to the observed period changes in many eclipsing systems. In the case of \n, the change of period is $\Delta P = 0.16$ sec only. 

 Many authors agree that, to explain the quasi-periodic oscillations of \oc\ values, it is necessary to combine several phenomena, i.e., not only the presence of additional bodies or the magnetic activity of components, but also a yet unknown phenomenon that is, for example, related to the internal structure of individual components 
\citep{ 2022MNRAS.513.2478M, 2022A&A...663A..90N, 2023ApJ...953...63H, 2025MNRAS.541.2553K, 2025MNRAS.544...24P}. 



\begin{figure}
\centering
\includegraphics[width=\columnwidth]{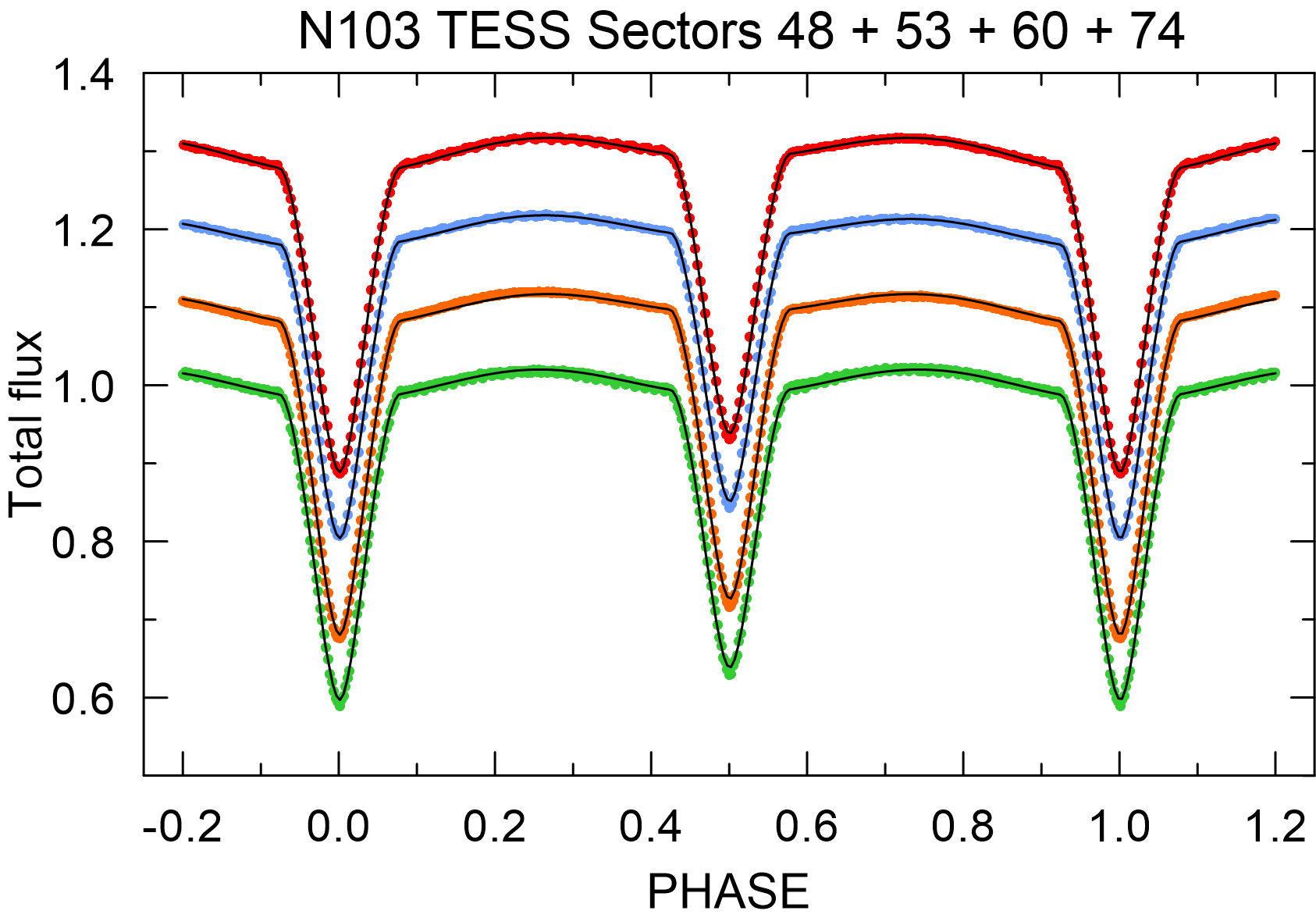}
\caption[]{{\sc Phoebe} solution of \n\ light curves: from up to bottom \T\ LC plots in Sectors 48 (red), 53 (blue), 60 (orange) and 74 (green) and their solution (black curves). 
Binning 300 was used. Individual curves are shifted by 0.1. 
Symmetric light curves without surface inhomogeneities are visible.}
\label{n103lc}
\includegraphics[width=0.45\textwidth]{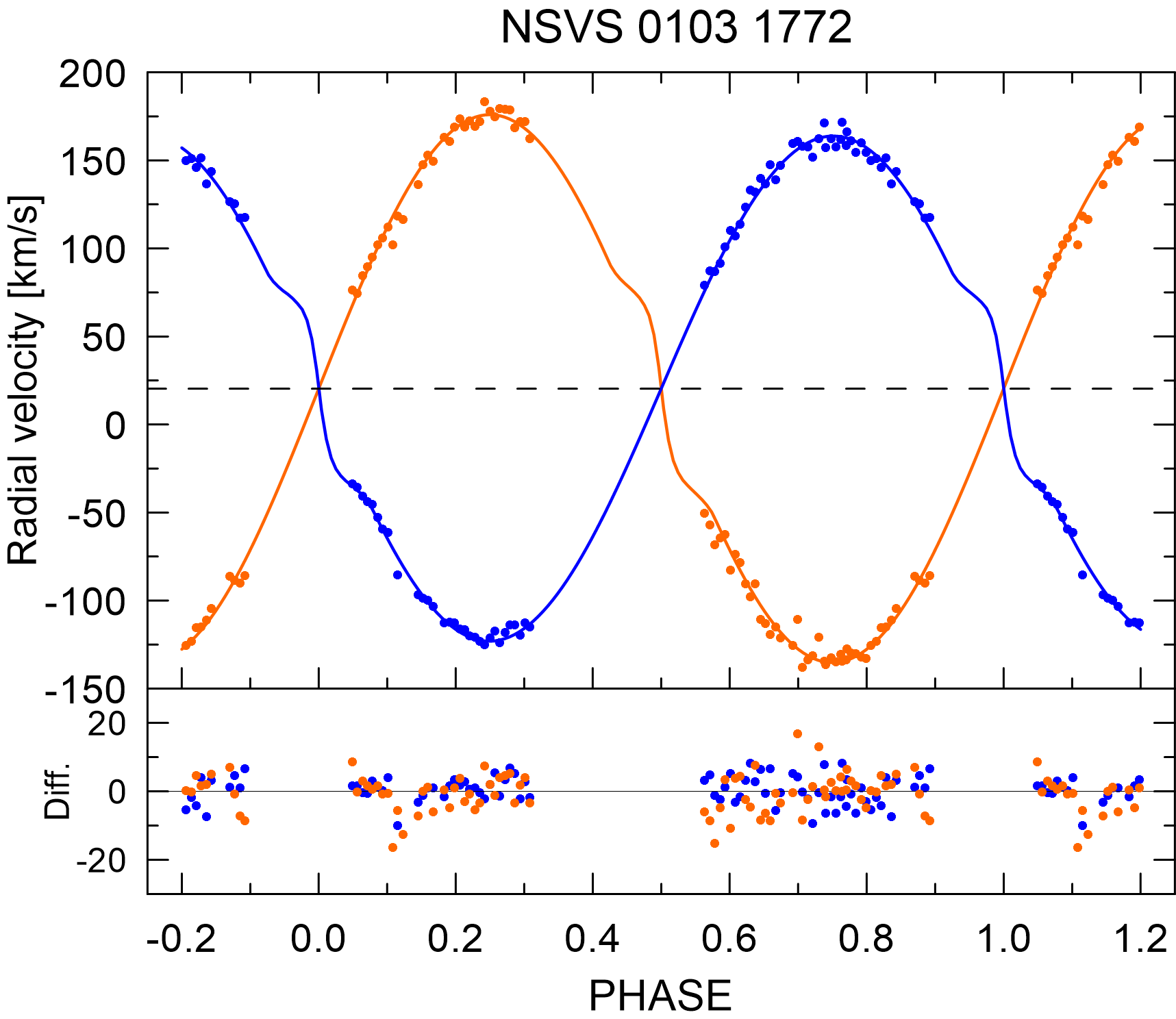}   
\caption[]{The radial velocity curve for \n\ obtained by \cite{2006astro.ph.10225L} and our solution in {\sc Phoebe}. Primary component as blue dots and curve, the secondary in orange. The $\gamma$-velocity (+20.4~km/s) is denoted as a dotted line.
Residuals are plotted in bottom panel. } 
\label{n103rv}
\end{figure}

\begin{figure}
\begin{center}
\includegraphics[width=\columnwidth]{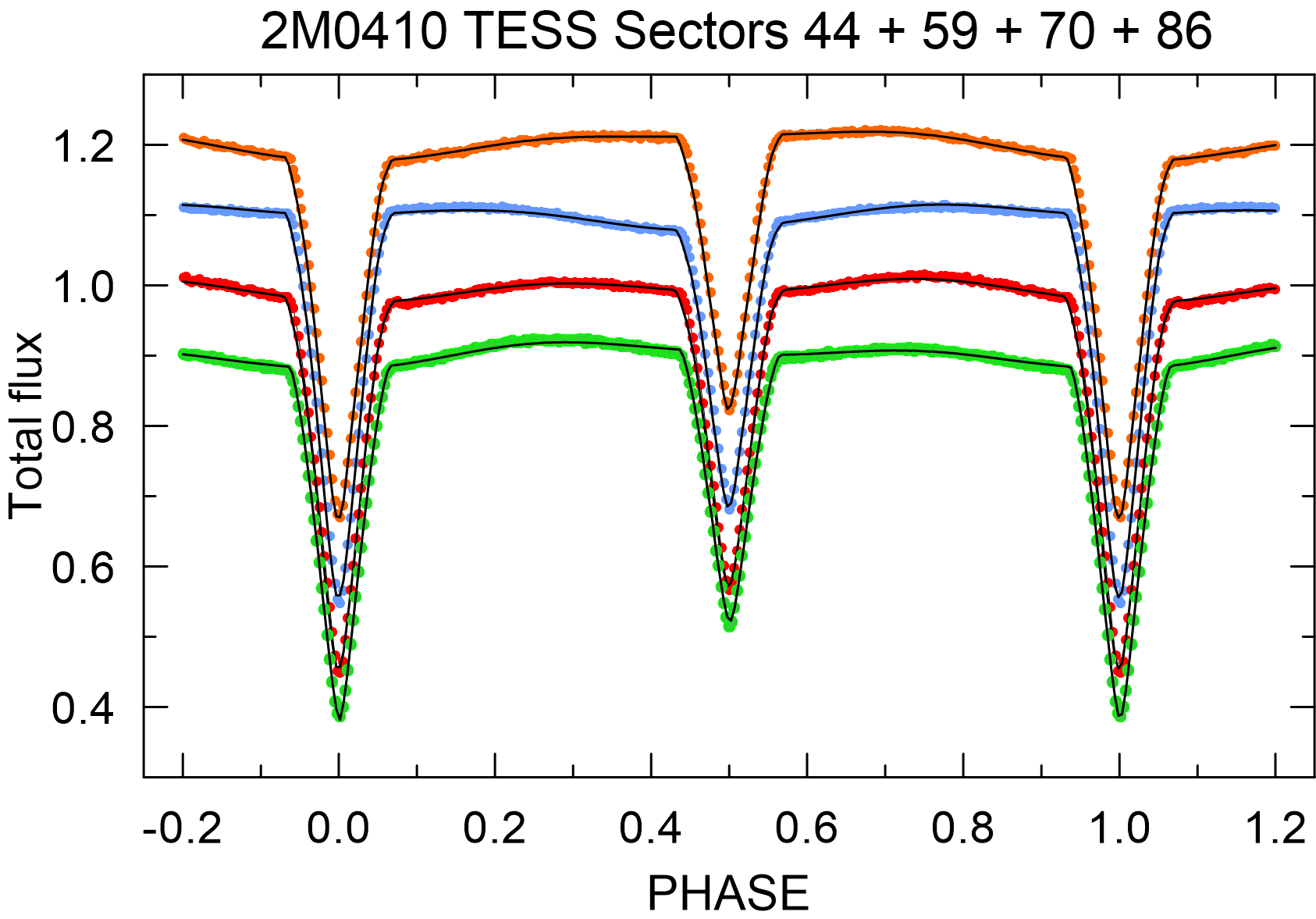}
\caption[ ]{ The {\sc Phoebe} final solution for the \T\ light curves of \obj, 
form top do bottom Sector~86 (orange), 70 (blue), 59 (red), and 44 (green) - shifted for clarity.  Binning of 300 points was applied. 
The resulting {\sc Phoebe} models as black curves. The out-of-eclipse changes on the light curve in individual sectors are clearly visible.} 
\label{2M04TESS}
\includegraphics[width=0.45\textwidth]{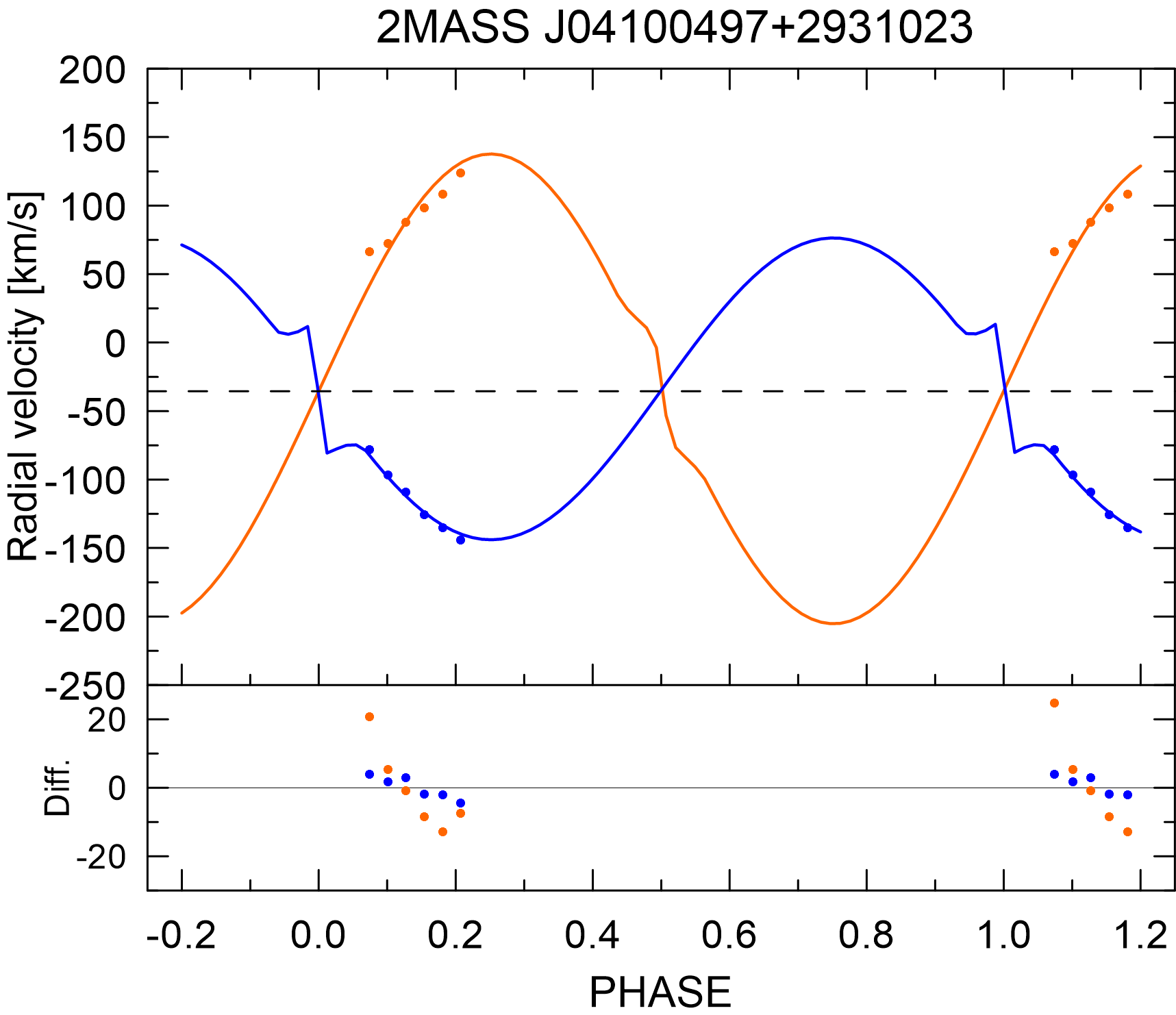}
\caption[ ]{ Partially covered RV curve of \obj\ obtained by \cite{2021RAA....21..115M}. 
The systemic $\gamma$-velocity (--33.8~km/s) is denoted as a dotted line.
Residuals are plotted in bottom panel. } 
\label{2M04RV}
\end{center}
\end{figure}

\section{ {\sc Phoebe} solution}  
\label{sec:curve}

As a first attempt, we selected four precise \T\ light curves for each binary to obtain the basic photometric parameters of the system, see Table~\ref{tess_sec}. 
These high-quality light curves were analyzed using the widespread {\sc Phoebe} code \citep{2005ApJ...628..426P, 2016ApJS..227...29P}, which is based on the Wilson-Devinney algorithm \citep{1971ApJ...166..605W} and is widely used to model the photometric light curves of eclipsing binaries as a standard tool. 
Because \n\ and \obj\ belong to late-type binaries, we adopted the bolometric albedos and gravity darkening coefficients as $A_1 = A_2 = 0.5$ and $g_1 = g_2 = 0.32$, which correspond to the convective envelopes (see \cite{1968ApJ...151.1123L}). Synchronous rotation was assumed for both components of the system ($F_1 = F_2 = 1$) and a circular orbit ($e = 0$). We used the linear cosine limb-darkening law with the coefficients adopted from \cite{1993AJ....106.2096V} tables.
 The radial velocities of \cite{2006astro.ph.10225L} and \cite{2021RAA....21..115M} were used as important input files. Accordingly, the temperature of the primary component was fixed to values obtained by previous investigators. 
The adjustable parameters were the mass ratio $q$, the {\bf semi-major axis} $a$, the systemic velocity $\gamma$, the inclination $i$, the effective temperature of the secondary component $T_2$, the luminosities $L_1$, $L_2$, and the dimensionless potentials of both components $\Omega_1, \Omega_2$. 
The fine and coarse grid raster for both components was set to 30.

In the case of \obj, the characteristics of the dark spot on the secondary component (colatitude, longitude, spot radius, and temperature factor) in each sector were taken into account (see Table~\ref{t8}).
The need to include a spot in the final solution was evident in view of the modulation of out-of-eclipse light curves (see Fig.~\ref{2M04TESS}).
Because the effective temperature and the radius of a spot are strongly correlated,  we assumed that the ratio of spot/star temperatures is close to 0.95 in our analyses.

Numerous {\sc Phoebe} runs in a detached mode using different setups of initial parameters were evaluated. The value of the cost function and the resulting parameters were recorded. The final solution was accepted when subsequent iterations did not result in a decrease in the {\sc Phoebe} cost function. 
The final solution of the \T\ light and radial velocity curves, compared to the previous results, is given in Tables~\ref{t3} and \ref{t4}, where the masses, radii, temperatures, potentials, and relative radii of both components are  presented. 
As one can see, the agreement between the previous and newly derived parameters 
 in the case of \n\ is relatively good within the given errors.
 Unfortunately, concerning the physical parameters of \obj, we cannot confirm the previous results of \cite{2021RAA....21..115M}, see Table~\ref{t4}. Our results do not support the more massive secondary compared to the primary component. 
 The parameters of the dark surface structure on the secondary component of \obj\ are given in Table~\ref{t8}.


\begin{table*}
\begin{center}
\caption{Comparison of absolute parameters of \n\ of previous authors with 
our {\sc Phoebe} solution of \T\ light curves. The mean values from four sectors are given.
Number of spots and their temperature factors used in the previous solution are listed.}
\label{t3}
\begin{tabular}{cccccccc}
\hline\hline
Parameter & Unit & Primary  & Secondary  &  Primary  &  Secondary & Primary  & Secondary  \\
          & & \multicolumn{2}{c}{\cite{2006astro.ph.10225L}} 
            & \multicolumn{2}{c}{\cite{2012IAUS..282..490W}}  
            & \multicolumn{2}{c}{this paper} \\
\hline\noalign{\smallskip}
$M_1, M_2$  & \ms & 0.5428(27) & 0.4982(25) & 0.555(5) & 0.495(5) & 0.5475(35) & 0.5038(40)  \\
$R_1, R_2$  & \rs & 0.5260(28) & 0.5088(30) & 0.592(5) & 0.492(5) & 0.5297(35) & 0.5217(35)  \\
$T_1, T_2$  & K  & 3615(72)  & 3513(31)   & 3500 (fixed) & 3455(25) & 3615(fixed) & 3580(50) \\
$L_1, L_2$  & \ls & 0.0426(34) & 0.0356(13) &   --  &   --    &  0.0423(25) & 0.0395(25)   \\
$M_{\rm bol1}, M_{\rm bol2}$ & mag & 8.08(0.25) & 8.27(0.12) & -- & -- & 8.15(0.30) & 8.26(0.25)  \\
$\Omega_1, \Omega_2$ & -- & 5.121(0.015) & 5.053(0.022)  &  -- &  -- & 5.08(0.12) & 5.01(0.15) \\

$q = M_2/M_1$ & -- &  \multicolumn{2}{c}{0.9217(0.0048)}
                 &  \multicolumn{2}{c}{0.891(0.0028) }
                 &  \multicolumn{2}{c}{0.9203(0.0028)}  \\ 

$a$  & \rs  & \multicolumn{2}{c}{2.189} & \multicolumn{2}{c}{--} & \multicolumn{2}{c}{2.190} \\

$i$  & deg      &  \multicolumn{2}{c}{85.91(0.03)}
                &  \multicolumn{2}{c}{85.7(0.1)  } 
                &  \multicolumn{2}{c}{83.58(0.1) } \\


No. of spots & -- & \multicolumn{2}{c} {2} & \multicolumn{2}{c}{1}  &  \multicolumn{2}{c}{0}   \\  
Temp. factor & -- & \multicolumn{2}{c}{1.048, 1.196} &  \multicolumn{2}{c}{0.95} &
 \multicolumn{2}{c}{--} \\
\hline
\end{tabular}
\end{center}
\end{table*}


\begin{table*}
\begin{center}
\caption{Comparison of absolute parameters of \obj\ of previous author with 
our {\sc Phoebe} solution of \T\ light curves. The mean values from four sectors are given.}
\label{t4}
\begin{tabular}{cccccccc}
\hline\hline
Parameter & Unit & Primary  & Secondary  &  Primary  & Secondary  \\
          & & \multicolumn{2}{c}{\cite{2021RAA....21..115M}} 
            & \multicolumn{2}{c}{this paper} \\
\hline\noalign{\smallskip}
$M_1, M_2$  & \ms & 0.587(6) & 1.049(11) & 0.639(45)  & 0.609(45)    \\
$R_1, R_2$  & \rs & 0.74(4)  & 0.68(4)   & 0.655(35)  & 0.631(35)    \\
$T_1, T_2$  & K   & 4500(6)  & 4209(161) & 4500(fixed) & 4200(200) \\
$M_{\rm bol1}, M_{\rm bol2}$ & mag & 8.08(0.25) & 8.25(0.12) & 7.08(0.10) & 7.72(0.12)  \\
$\Omega_1, \Omega_2$ & -- & 6.633(0.032) & 9.960(0.085)   &  6.375 & 6.485  \\
$q = M_2/M_1$ & -- &  \multicolumn{2}{c}{1.79(0.01)}
                   &  \multicolumn{2}{c}{0.95(0.12) } \\ 
$a$  & \rs  & \multicolumn{2}{c}{3.55} & \multicolumn{2}{c}{3.25}  \\
$i$  & deg      &  \multicolumn{2}{c}{89.1(0.2)}
                &  \multicolumn{2}{c}{87.9(0.5) }  \\
               
No. of spots & -- & \multicolumn{2}{c} {0}  & \multicolumn{2}{c}{1} \\
Temp. factor  & -- & \multicolumn{2}{c}{--}  & \multicolumn{2}{c}{0.95} \\
\hline
\end{tabular}
\end{center}
\end{table*}

\begin{table}
\begin{center}
\caption{Parameters of the dark surface structure on the secondary
 component of \obj. }
\label{t8}
\begin{tabular}{lcccccc}
\hline\hline
Parameter                & \multicolumn{4}{c}{\T\ Sector}   \\     
                         & 44    & 59    & 70 &   86  \\
\hline\noalign{\smallskip}
Colatitude [deg]        & 60     & 60    & 60   &  60  \\
Longitude [deg]         & 290    & 120   & 60   &  90  \\
Spot radius [deg]       & 15     & 15    & 15   &  15  \\
Temperature factor      & 0.93   & 0.95  & 0.94 &  0.96 \\
{\sc Phoebe} cost function & 350 &  510  &  785 &  670 \\      
\hline
\end{tabular}
\end{center}
\end{table}

\section{Flare activity}
\label{sec:flare}

\begin{figure}
\centering
\includegraphics[width=0.45\textwidth]{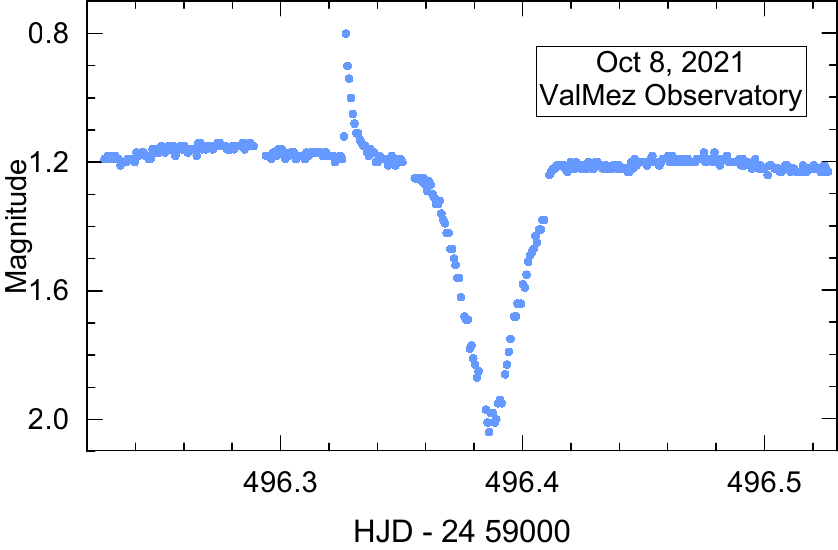}   
\caption[]{Example of one prominent eruption on \n\ followed by a primary eclipse which was discovered on the light curve obtained at \valmez\ observatory on October~8, 2021. The amplitude of eruption in filter $R$ was 0.41~mag, duration about 18~minutes. See also Table~\ref{eruptab}.} 
\label{n103erup}
\end{figure}

\begin{table}
\begin{center}
\caption{Number of flares on \n\ found in different TESS sectors. }
\label{eruptess}
\begin{tabular}{cccccc}
\hline\hline			
Sector & Duration & Number     & Frequency      \\
  No.  & [hours]  & of flares  & [flares/hour]   \\	
\hline \noalign{\smallskip}
14 & 621.576 &	12 & 0.0193 \\
20 & 594.744 &	15 & 0.0252 \\
21 & 633.936 &   6 & 0.0094 \\
26 & 574.656 &  13 & 0.0226 \\
40 & 654.504 &  13 & 0.0198 \\
41 & 611.232 &  14 & 0.0229 \\
47 & 595.92  &  18 & 0.0302 \\
48 & 481.296 &  12 & 0.0249 \\
53 & 577.608 &  11 & 0.0190 \\
60 & 417.95  &  11 & 0.0263 \\
73 & 386.30  &  6  & 0.0155 \\
74 & 447.36  &  9  & 0.0201 \\
75 & 407.11  & 17  & 0.0418 \\
\hline\noalign{\smallskip}
All sectors & 7004.19 & 157 & 0.0228 \\
\hline
\end{tabular}
\end{center}
\end{table}

Stellar flares are a typical manifestation of chromospheric activity, especially in red dwarf stars, when hot plasma is ejected into the surroundings and releases a quantum of accumulated magnetic energy. The total released energy
in the flares usually ranges from $10^{24} - 10^{27}$~J \citep{1989SoPh..121..299P}.
The significant eruption activity is well known in several low-mass eclipsing binaries: 
CM~Dra \citep{2007IBVS.5789....1N, 2009Ap.....52..512K, 2025MNRAS.541.2553K}, 
YY~Gem \citep{1990A&A...232...83D}, 
DV~Psc \citep{2010NewA...15..362Z}, 
NSVS~6550671 \citep{2010MNRAS.406.2559D}, 
CU~Cnc \citep{2012MNRAS.423.3646Q}, 
GJ~3236 \citep{2017MNRAS.466.2542S},
NSVS~7453183 \citep{2023MNRAS.520..353S},
 and recently in CC~Eri \citep{2026MNRAS.545f1993L}.

Stellar activity in a sample of active M~dwarfs was recently explored
by \cite{2025arXiv251002693R}. They found that the flare occurrence rate displays a flat distribution between the spectral types M0~--~M4 
($T_{\rm eff} \sim 3900 - 3200$ K) and decreases for later types. Faster rotators with $P_{\rm rot}$ < 1 day exhibit a higher flare occurrence rate and flare activity.  
M dwarfs with a higher flare occurrence rate tend to exhibit lower flare amplitudes, indicating that frequent flares in these M~dwarfs are generally less energetic.
Stellar flares are still at the forefront of interest and detailed research, including their connection with the existence of nearby exoplanets; see e.g. \cite{2024MNRAS.535.1000Y, 2024ARep...68..865Z}.

In our photometric campaign, \n\ was measured every clear night at several observatories
from 2013 to 2025 (see Section~\ref{sec:ground}). An example of a light curve with a prominent flare obtained in the $R$ filter is plotted in Fig.~\ref{n103erup}.
The eruption activity on \n\ was first presented by \cite{2016OEJV..175....1S}.
They recognized five individual flares observed during 2013–2015.
For the strongest eruption in March~2015, the total energy released in the two pass-bands $E_{\rm Rc} = 3.0 \times 10^{27}$~J in $R_c$ and $E_{\rm c} = 1.3 \times 10^{28}$~J in C was estimated, respectively.  Later, in a short report for the years 2013–2019, \cite{2020OEJV..208...29S} states the frequency 0.0331~eruptions per hour (or 1~eruption per 30.25~hours).

We used the same flare detection criterion as in our previous studies of the active eclipsing binary GJ~3236 \citep{2017MNRAS.466.2542S} or NSVS~7453183 \citep{2023MNRAS.520..353S}: a minimal difference of 0.005~mag from the surrounding continuum and at least three points on the descending curve.

The following results are based on the photometry available from the TESS satellite.  
N103 was monitored during 13~sectors (S14 - S75) from July 2019 to February 2024 for 7004 hours, see Table~\ref{tess_sec}. During this time interval, 157~flares were recorded. This gives a mean flare frequency of 0.0228 per hour (or 1 flare per 43.7 hours); see Table~\ref{eruptess}.
Similar results are based on our observations. In the period from July 2013 to June 2025, the binary was observed during 215~nights for a total of 1158~hours. During this time, 29~flares were recorded (frequency 0.025043, or 1~flare per 40~hours). Detailed information on several prominent flares is summarized in Table~\ref{eruptab}.

\begin{figure}
\centering
  \includegraphics[width=0.95\linewidth]{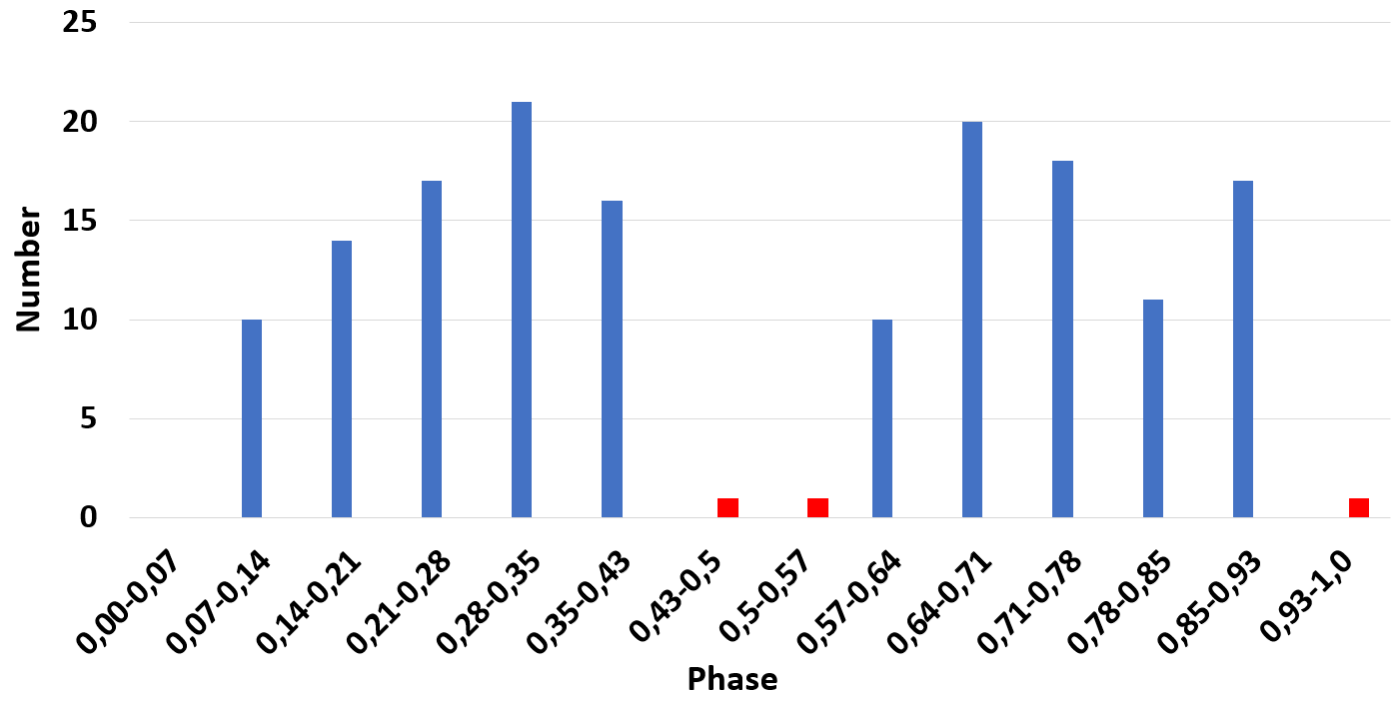}
  \includegraphics[width=0.95\linewidth]{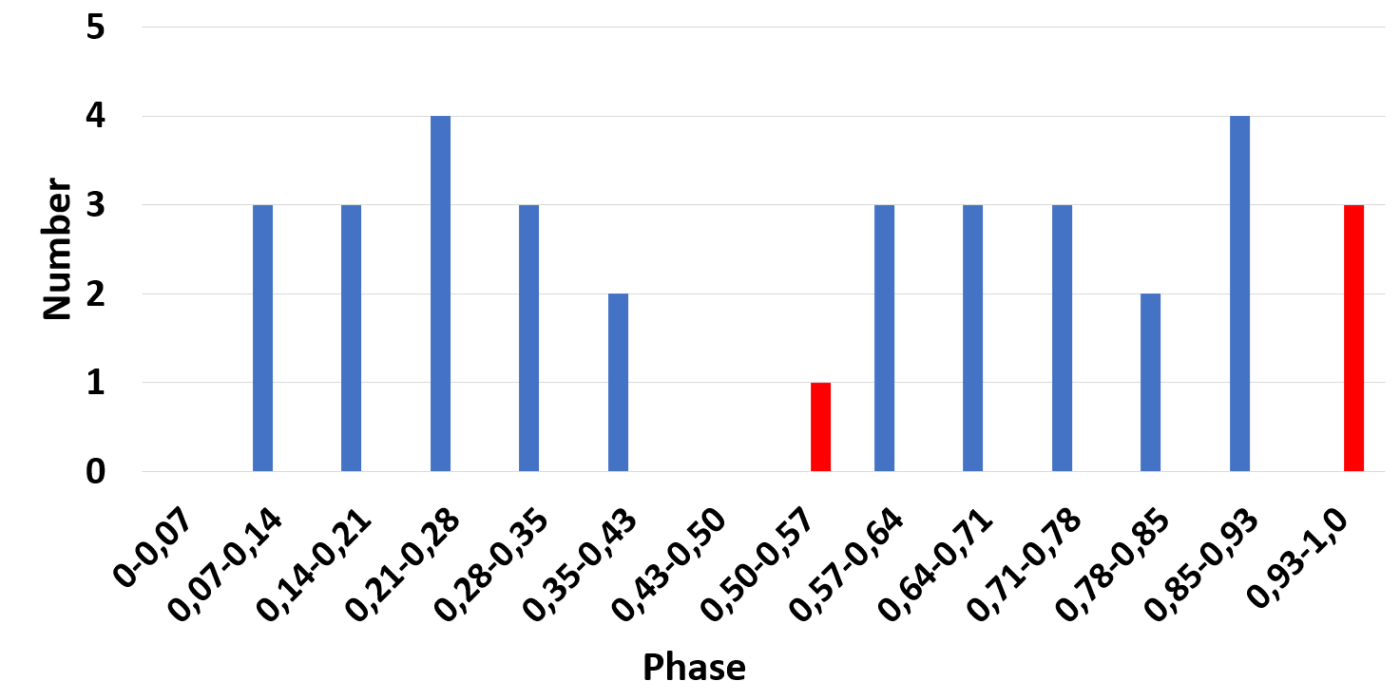} 
    \caption{Phase dependence of eruption activity of \n\ detected in 13~\T\ sectors during 2019-2024 (top panel) and at \valmez\ observatory during 2014-2025 (bottom). 
    Significantly fewer eruptions were recognized during both eclipses (red columns).
    Total number of eruptions is 157 and 34, respectively.}
    \label{n103erup2}
\end{figure}

\begin{table}
\begin{center}
\caption{Parameters of several prominent flares on \n\ observed at \valmez\ 
 observatory during 2015 - 2025. }
\label{eruptab}
\begin{tabular}{cccccc}
\hline\hline			
No. & Max JD Hel. & Event & Duration & Amplitude & Filter \\
	& -2400000    &       &	[min]    & [mag]     \\	
\hline\noalign{\smallskip}			
1 &	57089.26995 & Max & 103.0 &	0.35 & R \\				
2 &	57298.35013 & Max & 32.1 & 0.33 & V \\
3 & 57980.35750 & Max & 26.0 & 0.71 & V \\					
4 & 59161.21476 & Max & 19.5 & 0.45 & V \\
5 & 59060.49768 & Max & 11.5 & 0.32 & V \\
  &	   "	  &  "  &      & 0.10 & R \\
  &	   "	  &  "  &      & 0.03 & I \\
6 & 59496.32644 & Max & 17.9 & 0.41 & R \\
7 & 60389.56527 & Max & 21.9 & 0.17 & Clear \\
8 & 60808.53120 & Rise & 0.7 &  --  & -- \\
  & 60808.53168 & Max & 16.7 & 0.03 & Clear \\
\hline
\end{tabular}
\end{center}
\end{table}

The values of the eruption frequencies from different data sets indicate that the activity of this eclipsing binary is rather stochastic (random - regardless of the time range of the data series). 
Fig.~\ref{n103erup2} shows the phase dependence of eruption activity detected in \T\ sectors and at the \valmez\ observatory. The selected step in phase 0.07 corresponds to the duration of the eclipse. 
Surprisingly, shortly after the primary minimum (phases 0.0 -- 0.7), no eruption was registered in either data set. Moreover, a significantly lower number of eruptions during eclipses was identified in both data files (phases 0.93 -- 0.07, 0.43 -- 0.57).   
This may be partly due to the relatively rapid decrease/increase of the system's brightness during an eclipse ($\sim$~0.012 mag/min),  which may contribute to the poorer detectability of any flares.
Another possible explanation is that flares occur more frequently between the binary components on their facing hemispheres, which are less visible during both eclipses. It is therefore also possible that in N103 we see a result of magnetic interactions between the components. The corotating magnetic loops on both stars could be large enough to reach each other, and therefore the flux tubes can be temporarily bridged \citep{1980ApJ...239..911S, 2021ApJ...923...13C}. 
The distance between components in \n\ is about 1~\rs\ only. 
The role of inter-binary magnetic fields in flare activity is still not fully understood, and other phenomena (tidal forces or additional mutual heating of atmospheres) may play a role.

\section{Conclusions}
\label{sec:concl}

\begin{figure}
\centering
  \includegraphics[width=0.90\linewidth]{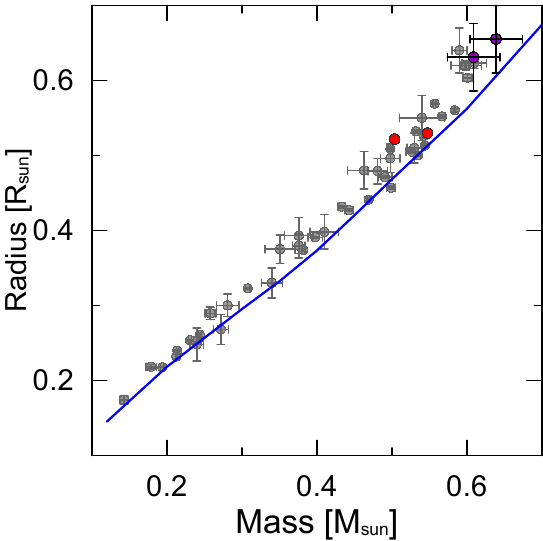}
    \caption[]{Mass-Radius diagram for known low-mass eclipsing binaries (collected by \cite{2018MNRAS.476.5253C}, gray dots). 
    The blue line is the 5~Gyr isochrone of the standard model from \cite{1998AA...337..403B}. 
    Two red dots denote the position of \n\ components, two magenta dots corresponds to \obj\ components.}
    \label{MRD}
\end{figure}

A study of late-type and low-mass binaries provides us with important information about the most frequent stars in our Galaxy. 
New precise {\sc Tess} photometry, supplemented by our long-term ground based photometric observations, was used to determine the photometric parameters of the components and improve their absolute parameters. 
The positions of \n\  and \obj\ components in the Mass-Radius diagram are plotted in Fig.~\ref{MRD}, where data for known low-mass eclipsing binaries were collected by \cite{2018MNRAS.476.5253C}.
The isochrone of 5~Gyr according to the standard models from \cite{1998AA...337..403B}, calculated for a solar metallicity ($[M/H]=0$), and the helium abundance of $Y=0.275$ is plotted. It is clear that the derived parameters of all components of  \n\ and \obj\ fit well among other low-mass systems with non-inflated radii compared to the 5~Gyr model. 

The high number of new mid-eclipse times was measured in the past, and the respective \oc\ diagrams were constructed. Eclipse timing variations have been used to infer the existence of stellar or sub-stellar companions. 
Concerning \n, the third body orbiting the eclipsing pair announced in our previous study \cite{2012IAUS..282..490W} was confirmed, and more precise LITE elements were obtained. Moreover, our dynamical tests indicate that \n\ cannot be a quadruple system. 
The detached eclipsing binary is orbited by a third body with a period of  19.4~years. It could be a brown dwarf with a minimal mass of $M_3 \simeq 50$~M$_{\rm Jup}$. The \oc\ diagram assembled in Fig.~\ref{n103} represents one of the most detailed descriptions of the period changes of a low-mass eclipsing binary. 
The short systematic deviation of secondary eclipse times between epochs 16~000 -- 17~000 ($\sim$ 150~days) could be caused by the presence of a temporary surface structure on one component.
 In the case of \obj, the third body was proposed in this system for the first time. 
The companion could be a red dwarf with a minimal mass of about 0.1~\ms\ in a circular orbit with a rather short period of about 2.1~years.
A certain similarity exists with another possibly quadruple low-mass eclipsing system NSVS~7453183 in the hierarchy (2+1)+1, where low eruption activity was also found \citep{2023MNRAS.520..353S}.

 Recently, \cite{2025MNRAS.544...24P} studied eclipse timing variations in seven well-known PCEBs, reviewing alternative mechanisms of \oc\ changes, including magnetic effects.  They conclude that although we cannot exclude the presence of circumbinary bodies, a combination of several other mechanisms may be required to explain the observed \oc\ diagrams. 
The similar result for white dwarf binaries was obtained by \cite{2026arXiv260217800Y}. They conclude that an Applegate- or Lanza-like magnetic mechanism is the most likely driving force for the timing variations seen
in the majority of these systems. In our case, \n\ is a pair of normal red dwarfs, where the observed period changes cannot be explained by a multiple system. 


 The statistics and characteristics of the flare events and dark regions on the surface of the components were estimated based on TESS and our own photometry. For N103, a mean frequency of flares of one per 40 hours was determined. In the case of 2M0410, practically no flares were detected.


Although the two systems discussed above are astrophysically very similar, in one case we observe smooth, symmetric light curves with a significant number of eruptions on the surface (\n), while the second system (\obj) has deformed light curves and some dark regions were introduced on the surface to solve their shape.
However, the eruptions are very sparse.
N103 is also a good example that not all sinusoidal variations of the period frequently visible on the \oc\ diagrams can be automatically attributed to a multiple (sub)stellar system.   

Long-term systematic monitoring of both of these objects in the future can provide more information about the nature of period changes and their surface activities. Complete coverage of the radial velocity curve of \obj\ would be suitable for the determination of accurate masses of the components.
 It is also a challenge for theoreticians to clarify the origin of triple systems with a giant planet or a brown dwarf as the distant orbiting body. 

\section*{Acknowledgments}

The useful suggestions and recommendations of an anonymous referee helped us improve the clarity and correctness of the paper and are greatly appreciated.
The research of MW and PZ was partially supported by the 
{\sc Cooperatio -- Physics} project of Charles University in Prague.
HK and KH were supported by the RVO project: 67985815.
The authors would also like to thank Lenka Kotkov\'a, \ond\ observatory, 
Marek Chrastina and Jan Janík, Masaryk University Brno,
Jan Vra\v{s}til, Tereza Je\v{r}\'abkov\'a, and Ond\v{r}ej Chrenko, 
all former students of Charles University in Prague, for their important contributions to photometric observations during the past decades. 
The authors thank Dr. M. Lopez-Morales for the use of their original radial velocities.
TH thanks Jan Zahajsk\'y (Supra Praha) and Martin Myslivec, who played a significant role in putting the whole set in the Dark Sky Beskydy observatory into operation.
This paper includes data collected by the {\sc Tess} mission.
Funding for the {\sc Tess} mission is provided by the NASA Science Mission directorate. 
Some of the data presented in this paper were obtained from the Mikulski Archive for Space Telescopes (MAST).
This work has used data from the European Space Agency (ESA) mission {\sc Gaia}~\footnote {{\sc Gaia}, \url{https://www.cosmos.esa.int/gaia}}, processed by the {\sc Gaia} Data Processing and Analysis Consortium 
(DPAC)~\footnote{{\sc Dpac}, \url{https://www. cosmos.esa.int/web/gaia/dpac/consortium}}. 
Funding for the DPAC has been provided by national institutions, in particular
the institutions participating in the {\sc Gaia} Multilateral Agreement. 
This research has also used the Keck Observatory Archive (KOA), 
which is operated by the W.M. Keck Observatory and NASA Exoplanet Science Institute (NExScI), under contract with the National Aeronautics and Space Administration.
The following Internet-based resources were used in the research for this paper:
the SIMBAD database \citep{2000A&AS..143....9W} 
and the VizieR catalog access tool \citep{2000A&AS..143...23O}
operated at CDS, Strasbourg Astronomical Observatory, France, 
the NASA's Astrophysics Data System Bibliographic Services, 
and VarAstro of the Czech Astronomical Society.
This investigation is part of an ongoing collaboration 
between professional astronomers and the Czech Astronomical Society, 
Variable Star and Exoplanet Section.

\section*{Data Availability}

Some of the data were derived from sources in the public domain, 
and the respective URLs are provided as footnotes. 
The other data are available on reasonable request from the authors.

\printcredits


\bibliographystyle{cas-model2-names}

\bibliography{example}


\appendix

\section{Tables}

\begin{table*}
\centering
\small 
\caption{New ground-based times of primary and secondary eclipses of \n\ and \obj\ obtained at different observatories since 2007. The complete list of 443 mid-eclipse times of \n\ is given in the electronic part of this work.  }
\label{t1}
\begin{tabular}{llccc}
\hline\hline
      BJD & Error  & Filters & Epoch & Observatory  \\
-24 00000  & [day] &         &       &  \\        
\hline
 54266.40742 & 0.0001 & $R$ & 2199.5 &  \ond \\
 54278.37212 & 0.0001 & $R$ & 2232.0 &  \ond \\
 54532.38946 & 0.0002 & $R$ & 2922.0 &  \ond \\ 
 54581.35187 & 0.0001 & $R$ & 3055.0 &  \ond \\
 54840.52308 & 0.0001 & $VR$ & 3759.0 &  \ond  \\
 54840.70663 & 0.0001 & $VR$ & 3759.5 &  \ond \\
 55143.3178* & 0.0002 & $BVR$ & 4581.5 &  Mt.Suhora \\
 55186.39034 & 0.0002 & $R$   & 4698.5 & MUO  \\ 
 55192.28084* & 0.0001 & $VR$ & 4714.5 & MUO  \\
 55192.46471* & 0.0001 & $VR$ & 4715.0 & MUO  \\
 ...         &  ...   & ...  & ...    & ...  \\
     61165.53559 & 0.0001 & $g'r'$ & 20940.0 & \valmez \\
     61166.45591 & 0.0001 & $g'r'$ & 20942.5 & \valmez \\
     61170.50550 & 0.0001 & $g'r'$ & 20953.5 & \valmez \\
     61182.46990 & 0.0002 & $g'r'$ & 20986.0 & \valmez \\
\hline
  59172.29167* & 0.0001 & $VR$ &   538.0 & \ond \\
  59250.39919* & 0.0004 & $VR$ &   666.5 & \ond \\
  59268.33059* & 0.0001 & $VR$ &   696.0 & \ond \\
  59581.37264* & 0.0001 & $VRI$ &  1211.0 & \ond \\
  59897.45312  & 0.0003 & $C$ &   1731.0 & \valmez \\
  59927.23790  & 0.0003 & $C$ &   1780.0 & \valmez \\
  59927.54197  & 0.0003 & $C$ &   1780.5 & \valmez \\
  59982.24754* & 0.0001 & $VRI$ & 1870.5 & \ond \\
  59998.35532  & 0.0003 & $C$ &  1897.0 & \valmez\\
  60005.34538  & 0.0002 & $C$ &  1908.5 & \valmez  \\
  60006.25722* & 0.0001 & $RI$ &  1910.0 & \ond \\
  60015.37473* & 0.0001 & $VR$ &  1925.0 & \ond \\
  60209.58177  & 0.0002 & $L$  &  2244.5 & Beskydy \\
  60235.41576  & 0.0002 & $C$ & 2287.0   &  \valmez  \\
  60241.49433  & 0.0002 & $C$ & 2297.0   &  \valmez  \\    
  60243.31769  & 0.0002 & $C$ & 2300.0   &  \valmez  \\   
  60261.24909  & 0.0003 & $C$ & 2329.5   &  \valmez  \\ 
  60290.42611* & 0.0002 & $VR$ & 2377.5  &  \ond  \\
  60297.41657  & 0.0001 & $R$ & 2389.0   &  \ond \\
  60297.41673  & 0.0002 & $C$ & 2389.0   &  \valmez  \\   
  60329.32874  & 0.0001 & $R$ & 2441.5   &  \ond  \\ 
  60340.26992  & 0.0003 & $C$ & 2459.5   &  \valmez  \\   
  60388.29014  & 0.0003 & $C$ & 2538.5   &  \valmez  \\   
  60412.30014  & 0.0002 & $VR$ & 2578.0  &  \ond \\ 
  60618.36044* & 0.0001 & $VR$ & 2917.0  &  \ond \\
  60624.43886  & 0.0001 & $C$ & 2927.0   &  \hol \\
  60674.28206* & 0.0001 & $VR$ & 3009.0  &   \ond  \\
  60688.26280  & 0.0005 & $C$ & 3032.0   &   \hol \\
  60688.26277  & 0.0001 & $R$ & 3032.0   &   \hol \\
  60708.32157* & 0.0001 & $VR$ & 3065.0  &   \ond \\
  60710.44837  & 0.0004 & $C$ & 3068.5   &   \hol \\
  60711.36073  & 0.0002 & $C$ & 3070.0   &   \hol \\
  60739.32165  & 0.0001 & $L$ & 3116.0   &   Beskydy \\
  60756.34111* & 0.0001 & $VR$ & 3144.0  &   \ond  \\
  60930.48868  & 0.0001 & $L$ & 3430.5   &   Beskydy \\
  61045.37219  & 0.0001 & $C$ & 3619.5   &   \valmez \\
  61058.44105* & 0.0001 & $VR$ & 3641.0  &   \ond \\        
  61058.44095  & 0.0001 & $C$  & 3641.0  &   \valmez \\ 
  61097.34337* & 0.0001 & $RI$ & 3705.0  &   \ond \\
\hline
\end{tabular}

Note: * mean value of VR, VI, RI, VRI, or g'r' measurements
\end{table*}

\begin{table*}
\caption{New times of primary and secondary eclipses of \n\ based 
on the \T\ photometry in different sectors.}
\label{tess}
\begin{center}
\begin{tabular}{cccccc}
\hline\hline
BJD           &  Epoch  & TESS   & BJD         & Epoch  &   TESS \\
-24 00000     &         & Sector & -24 00000   &          & Sector \\    
\hline
 58683.53465 & 14198.0  & 14 & 59579.95483 & 16633.0  & 47 \\ 
 58683.71842 & 14198.5  & 14 & 59580.13868 & 16633.5  & 47 \\ 
 58683.90273 & 14199.0  & 14 & 59580.32297 & 16634.0  & 47 \\ 
 58842.57099 & 14630.0  & 20 & 59580.50682 & 16634.5  & 47 \\ 
 58843.85929 & 14633.5  & 20 & 59744.14521 & 17079.0  & 53 \\ 
 58844.04354 & 14634.0  & 20 & 59744.32920 & 17079.5  & 53 \\ 
 58870.54954 & 14706.0  & 21 & 59744.51333 & 17080.0  & 53 \\ 
 58870.73352 & 14706.5  & 21 & 59744.69734 & 17080.5  & 53 \\ 
 58882.69820 & 14739.0  & 21 & 59937.05053 & 17603.0  & 60 \\ 
 58882.88212 & 14739.5  & 21 & 59937.23467 & 17603.5  & 60 \\ 
 58891.16536 & 14762.0  & 21 & 59938.15497 & 17606.0  & 60 \\ 
 58891.34933 & 14762.5  & 21 & 59938.52309 & 17607.0  & 60 \\ 
 59010.44266 & 15086.0  & 26 & 59938.70723 & 17607.5  & 60 \\ 
 59010.62654 & 15086.5  & 26 & 59947.17455 & 17630.5  & 60 \\ 
 59010.81078 & 15087.0  & 26 & 59947.35842 & 17631.0  & 60 \\ 
 59010.99467 & 15087.5  & 26 & 60286.04752 & 18551.0  & 73 \\ 
 59390.73089 & 16119.0  & 40 & 60286.23162 & 18551.5  & 73 \\ 
 59390.91472 & 16119.5  & 40 & 60286.41565 & 18552.0  & 73 \\ 
 59400.11815 & 16144.5  & 40 & 60312.92182 & 18624.0  & 74 \\ 
 59400.30261 & 16145.0  & 40 & 60313.10590 & 18624.5  & 74 \\ 
 59412.63492 & 16178.5  & 40 & 60317.33954 & 18636.0  & 74 \\ 
 59412.81913 & 16179.0  & 40 & 60317.52357 & 18636.5  & 74 \\ 
 59417.23696 & 16191.0  & 40 & 60335.01024 & 18684.0  & 74 \\ 
 59417.42072 & 16191.5  & 40 & 60335.93058 & 18686.5  & 74 \\ 
 59420.18217 & 16199.0  & 41 & 60339.98008 & 18697.5  & 75 \\  
 59420.36582 & 16199.5  & 41 & 60340.16421 & 18698.0  & 75 \\  
 59425.88796 & 16214.5  & 41 & 60340.34819 & 18698.5  & 75 \\  
 59426.07234 & 16215.0  & 41 & 60366.85438 & 18770.5  & 75 \\  
 59434.35513 & 16237.5  & 41 & 60367.03838 & 18771.0  & 75 \\  
 59434.53956 & 16238.0  & 41 \\   
 59440.61357 & 16254.5  & 41 \\  
 59440.79795 & 16255.0  & 41 \\
\hline
\end{tabular}
\end{center}
\end{table*}

\begin{table*}
\begin{center}
\caption{ New \T\ times of primary and secondary eclipses of \obj\
based on the \T\ photometry in five different sectors.}
\label{2m04mintess}
\begin{tabular}{cccc}
\hline\hline
BJD --     & Epoch &  BJD --     & Epoch   \\
24 00000   &       &  24 00000   &       \\
\hline
 \multicolumn{2}{c}{Sector 43} & \multicolumn{2}{c}{Sector 59}    \\  
  59474.39089 & 1035.0 &         59912.64892   &   1756.0   \\
  59475.30275 & 1036.5 &         59913.25718   &   1757.0   \\  
  59475.60653 & 1037.0 &         59914.47283   &   1759.0   \\    
  59476.21438 & 1038.0 &         59915.08052   &   1760.0   \\    
  59476.82229 & 1039.0 &         59915.38475   &   1760.5   \\   
  59477.12640 & 1039.5 &         59915.68850   &   1761.0   \\    
  59477.43026 & 1040.0 &         59916.29638   &   1762.0   \\    
  59478.64580 & 1042.0 &         59916.60025   &   1762.5   \\    
\multicolumn{2}{c}{Sector 44} &  59918.11988   &   1765.0   \\    
  59500.52857 & 1078.0 &         59918.72767   &   1766.0   \\   
  59501.13630 & 1079.0 &         59926.02211   &   1778.0   \\    
  59502.35182 & 1081.0 &         59928.45345   &   1782.0   \\    
  59503.87112 & 1083.5 &         59929.06131   &   1783.0   \\   
  59504.47880 & 1084.5 &         59929.66910   &   1784.0   \\  
  59505.39123 & 1086.0 &         59929.97320   &   1784.5   \\  
  59506.30248 & 1087.5 &         59930.88470   &   1786.0   \\ 
  59507.82252 & 1090.0 &         59931.49251   &   1787.0   \\ 
  59517.54827 & 1106.0 &         59932.10049   &   1788.0    \\ 
  59519.37192 & 1109.0 &         59933.31622   &   1790.0    \\ 
  59520.58743 & 1111.0 &         59933.62030   &   1790.5    \\ 
  59521.19544 & 1112.0 &         59933.92410   &   1791.0    \\  
  59522.10700 & 1113.5 &         59934.22821   &   1791.5    \\  
  59522.71512 & 1114.5 &         59934.53193   &   1792.0    \\
  59523.01888 & 1115.0 &       \multicolumn{2}{c}{Sector 80} \\
  59523.62672 & 1116.0 &       60636.29211 & 2946.5 \\
  59524.23471 & 1117.0 &       60637.20347 & 2948.0 \\ 
\multicolumn{2}{c}{Sector 70}& 60639.93870 & 2952.5 \\ 
  60209.27804 & 2244.0 &       60640.85069 & 2954.0 \\  
  60213.53301 & 2251.0 &       60641.45848 & 2955.0 \\   
  60213.83687 & 2251.5 &       60642.06627 & 2956.0  \\  
  60214.14087 & 2252.0 &       60643.28208 & 2958.0  \\  
  60215.35651 & 2254.0 &       60643.58587 & 2958.5  \\  
  60217.18034 & 2257.0 &       60643.89003 & 2959.0  \\  
  60218.39589 & 2259.0 &       60649.05639 & 2967.5  \\  
  60219.00381 & 2260.0 &       60657.26262 & 2981.0   \\ 
  60219.61148 & 2261.0 &       60657.87037 & 2982.0   \\ 
  60224.77806 & 2269.5 &       60658.47808 & 2983.0  \\  
  60225.08226 & 2270.0 &       60660.30180 & 2986.0  \\  
  60226.29795 & 2272.0 &       60662.73320 & 2990.0  \\   
\hline\noalign{\smallskip}
\end{tabular}
\end{center}
\end{table*}

\label{lastpage}
\end{document}